\documentclass[a4paper, 12 pt, twocolumn, final, conference]{ieeetran}
\IEEEoverridecommandlockouts

\usepackage{graphicx}
\usepackage{subcaption}
\usepackage{times}   
\usepackage{amsmath} 
\usepackage{amssymb} 
\usepackage{amsfonts}
\usepackage{algorithm}
\usepackage{algorithm, algorithmic}
\usepackage{array}
\usepackage{multirow}
\usepackage{microtype}
\usepackage{graphics}
\usepackage{booktabs} 
\usepackage{url}
\usepackage{bm}  
\usepackage{stfloats} 

\usepackage[colorinlistoftodos,prependcaption,textsize=tiny]{todonotes}

\usepackage[square, numbers, compress]{natbib}

\usepackage{color}

\usepackage[UKenglish]{isodate}

\usepackage{booktabs}
\usepackage[normalem]{ulem}
\useunder{\uline}{\ul}{}

\newcolumntype{P}[1]{>{\centering\arraybackslash}p{#1}}
\usepackage{array}

\title{\LARGE \bf
Deep Inception Networks: A General End-to-End Framework for Multi-asset Quantitative Strategies
}

\author{
    \IEEEauthorblockN{Tom Liu\IEEEauthorrefmark{1},  Stephen Roberts\IEEEauthorrefmark{2},
    Stefan Zohren\IEEEauthorrefmark{2}}
    \IEEEauthorblockA{\IEEEauthorrefmark{1}University of Oxford, \IEEEauthorrefmark{2}Oxford-Man Institute
of Quantitative Finance, University of Oxford}
}

\makeatletter
\renewcommand*{\thetable}{\arabic{table}}
\renewcommand*{\thefigure}{\arabic{figure}}
\let\c@table\c@figure
\makeatother 

\usepackage{caption}
\begin{document}
\maketitle
\thispagestyle{plain}
\pagestyle{plain}
%

\begin{abstract}
We introduce Deep Inception Networks (DINs), a family of Deep Learning models that provide a general framework for end-to-end systematic trading strategies. DINs extract time series (TS) and cross sectional (CS) features directly from daily price returns. This removes the need for handcrafted features, and allows the model to learn from TS and CS information simultaneously. 
DINs benefit from a fully data-driven approach to feature extraction, whilst avoiding overfitting. Extending prior work on Deep Momentum Networks, DIN models directly output position sizes that optimise Sharpe ratio, but for the entire portfolio instead of individual assets. We propose a novel loss term to balance turnover regularisation against increased systemic risk from high correlation to the overall market. Using futures data, we show that DIN models outperform traditional TS and CS benchmarks, are robust to a range of transaction costs and perform consistently across random seeds.  To balance the general nature of DIN models, we provide examples of how attention and Variable Selection Networks can aid the interpretability of investment decisions. These model-specific methods are particularly useful when the dimensionality of the input is high and variable importance fluctuates dynamically over time. Finally, we compare the performance of DIN models on other asset classes, and show how the space of potential features can be customised. 

\end{abstract}

\section{Introduction}

Many quantitative trading strategies are either based on theoretical (theory-driven) or empirical (data-driven) analysis. Theory-driven strategies attempt to explain patterns in market data with various rationales, ranging from economic to behavioural science. Common types include momentum \cite{moskowitz2012time,jegadeesh1993returns}, mean-reversion \cite{lo2011non}, technical sentiment \cite{bekaert2014vix}, value \cite{sp500_value}, growth \cite{chan2004valuegrowth} and quality \cite{ung2014quality}. Despite a limited number of strategy types, there can be considerable variation due to different timescales, feature selection, thresholds and other rules for signal creation. The style of implementation, time series (TS) or cross sectional (CS), also tends to differ by strategy and asset class \cite{baz2015dissecting}. For example, traditional strategies have traded momentum by considering the performance of assets independently (TS) \cite{moskowitz2012time}, or relative performance across assets (CS) \cite{jegadeesh1993returns}.

However, there is still a limit to the orthogonality of theory-driven strategies. Absolute profitability can also decay due to competition. To improve their alpha, researchers may turn to data-driven strategies that exploit complex market behaviours without needing to understand why they occur. Rather than handcrafting complex decision trees, which may fail to generalise on noisy and non-stationary financial data, a popular option in recent years has been to use Machine Learning (ML) models. ML models vary from predicting future price trends \cite{kim2019enhancing, abe2020cross} to more robust methods of predicting relative performance of assets \cite{poh2021building, song2017stock}. These ML-based strategies can improve risk-adjusted returns compared to traditional theory-driven strategies, prior to transaction costs. However, the position sizing rule often remains a separate step, resulting in ML models that only account for prediction error, but not risk or costs.

\begin{figure*}[!htbp]
  \includegraphics[width=0.9\textwidth]{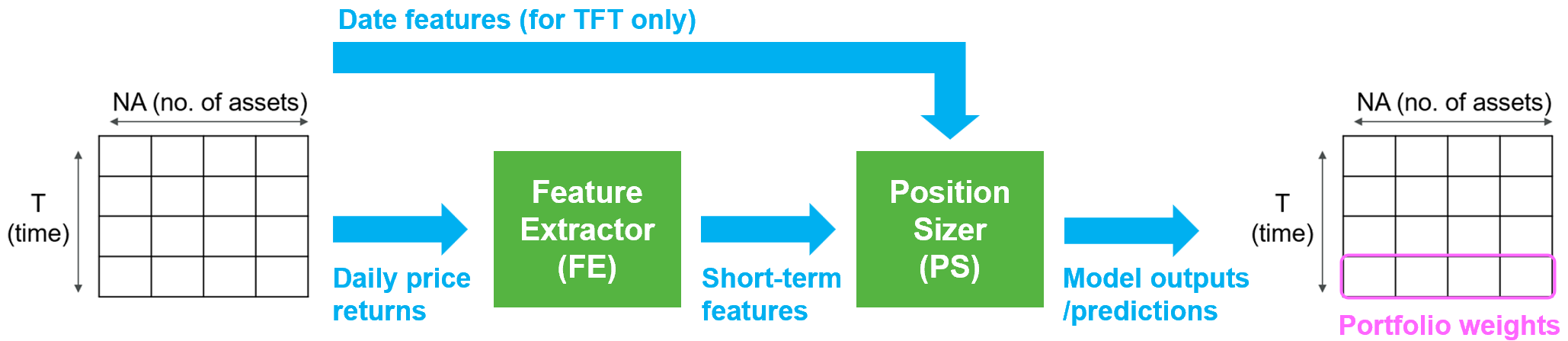}
  \caption{High-level model architecture for Deep Inception Networks (DINs).}
  \label{fig:ModelArchitecture_HighLevel}
\end{figure*}

A recent development are Deep Momentum Networks (DMNs) \cite{lim2019enhancing, wood2021trading, wood2022slow} that incorporate position-sizing explicitly into momentum strategies. DMNs output position sizes to directly optimise a risk-adjusted metric, such as Sharpe ratio. The loss function can also be modified to account for transaction costs. In addition to these single-asset TS models, \cite{tan2023spatio} considers CS features over time and uses multitask learning to output position sizes for all assets. However, this still requires the researcher to select input features, such as price trends over a range of timescales or crossover indicators. Optimising (a weighted combination of) the individual Sharpe ratios for each asset may also yield different results to optimising the overall portfolio Sharpe ratio.

Our proposed Deep Inception Networks (DINs) provide a general framework for an end-to-end strategy: the inputs are minimally processed data and the outputs are risk and cost-adjusted position sizes. The researcher only needs to choose what type of input data is needed, for example price returns, and the types of behaviour that the model should capture. The high-level model architecture for DINs is shown in Exhibit \ref{fig:ModelArchitecture_HighLevel}. First, we use a Feature Extractor (FE) to learn short-term features. This is followed by a Position Sizer (PS) that captures longer term dependencies, similar to \cite{lim2019enhancing, wood2021trading, wood2022slow, lim2021temporal}. Unlike these prior works, DINs create outputs for all assets simultaneously, optimising portfolio Sharpe ratio and taking advantage of CS information.

The DIN framework is inherently customisable, by choosing from a range of FE variants, which are designed to extract specific types of behaviour. Some FEs are based on prior work in TS prediction, for example DeepLOB \cite{zhang2019deeplob} and AxialLOB \cite{kisiel2022axial}. We also introduce novel FEs, Original Custom Inception Module (OrigCIM) and Flexible Custom Inception Module (FlexCIM): these are capable of learning CS and more complicated behaviour, such as lead-lag. The choice of FE can be determined systematically by comparing backtest results for the chosen assets. Critically, the features learnt by DIN models have semantic meaning. By constraining the possible feature ``shapes", DINs have less potential for overfit compared to universal end-to-end models, such as \cite{zhang2021universalE2E}, that can learn features from any combination of elements in the input matrix. 

The optimal PS variant can also be determined systematically. In this paper, we consider Long Short-Term Memory (LSTM) units \cite{lim2019enhancing} and Temporal Fusion Transformers (TFTs) \cite{wood2021trading,wood2022slow,lim2021temporal}. Both components enable the DIN to capture longer-term variations in the data. The key difference is that TFTs are able to recover ``forgotten" information from past cell states, using an attention layer. TFTs have been shown to yield superior results compared to LSTMs in \cite{wood2021trading,wood2022slow} and are also more interpretable. 

Evaluating on cross-asset futures data from 2005-2022, DIN models (OrigCIM x TFT) outperform all benchmarks before and after transaction costs. In particular, there is a \textbf{10x} increase in Sharpe ratio to \textbf{2.95}, compared to an equal-weight Long-only portfolio, prior to transaction costs. This is profitable until \textbf{4.82} basis points (bps) costs under a proportional cost model. To verify the potential for generalisation, we test the viability of DIN models on other asset classes: foreign exchange (FX), equities and cryptocurrencies. We find that the optimal choice of FE depends on the asset class. For futures and FX, where there are clusters of correlated assets, CS information captured by OrigCIM and FlexCIM is useful. For equities and cryptocurrencies, there is more homogeneous correlation structure: DeepLOB, which only extracts TS features, is best. Although we present results for daily trading, DeepLOB was originally developed for high-frequency Limit Order Book (LOB) applications, and DIN models could be similarly adapted to work at different frequencies by choosing a suitable FE module.

As a counterbalance to the general nature of DINs, it is important to understand the investment decisions output by the model. Model-agnostic LIME \cite{ribeiro2016LIME} and SHAP \cite{lundberg2017SHAP} methods are unsuitable due to the high dimensionality of the input matrix, and fluctuations in feature importance over time. Instead we focus on model-specific methods that enable local explanations. In particular, there are two subsets of neural network weights that are most insightful: Variable Selection Networks (VSNs) within the FlexCIM feature extractor and attention within the TFT position sizer. VSN weights show how the importance of different types of extracted features varies over time. The attention map can be used to identify which past timepoints have the largest effect on the ``one-step-ahead" prediction. We present case studies, such as the Great Financial Crisis (GFC) and COVID-19. VSN weights indicate salient trends in feature importance surrounding catalyst events, and attention focuses on historically similar time points rather than the most recent data.

We summarise key attributes of DINs below.
\begin{enumerate}
    \item \textbf{No hand-crafted features}: this is a fully data-driven architecture for both feature extraction and position sizing.
    \item \textbf{Balances generality against overfitting}: by selecting the FE module, we restrict the space of learnable features.
    \item \textbf{Interpretable decisions}: VSN and attention weights aid the understanding of model behaviour over time.
    \item \textbf{Optimises portfolio Sharpe}: this helps avoid concentrations in risk, which are not considered when optimising allocations in isolation.
    \item \textbf{Turnover and correlation regularisation}: improves robustness to costs, whilst avoiding overfit to strategies with static allocations. These may not generalise to unseen data.
\end{enumerate}

\section{Benchmark Strategies}
\subsection{Benchmarks}
\label{section:Methods_Benchmarks}
To assess the performance of DIN models, we select benchmarks from prior work on momentum strategies that capture TS or CS behaviour separately. 
\begin{itemize}
    \item \textbf{JT}: based on the seminal CS strategy in \cite{jegadeesh1993returns}.
    \item \textbf{LM}: a CS strategy using Learning to Rank models, specifically LambdaMART \cite{poh2021building}.
    \item \textbf{MOP}: based on the seminal TS strategy in \cite{moskowitz2012time}.
    \item \textbf{BAZ}: a TS strategy with MACD indicators \cite{baz2015dissecting}.
\end{itemize}

For each momentum strategy, we select either a long-only (LO) or long-short (LS) portfolio, based on $Sharpe$. The inputs to these strategies are heuristic features computed from price or returns data. We contrast this with a key advantage of DIN models, which may automatically learn a wider range of features that the researcher has not explicitly defined, such as reversion, spread or lead-lag. We also create a combined (CMB) portfolio of momentum benchmarks and compare this with the inherent ability of DINs to combine CS and TS information.
\begin{itemize}
    \item \textbf{CMB EW}: an equal-weight combination.
    \item \textbf{CMB VS}: a volatility scaled combination.
\end{itemize}

The final benchmark is \textbf{Long-only}, which takes equal positive weights for all assets. Further details on all benchmarks are presented in Appendix \ref{apdx:benchmarks}.

\subsection{Portfolio Construction}
\label{section:Methods_Portfolio}
Each benchmark strategy or DIN model produces outputs which are portfolio weights $w_{i,t}$ for all assets $i$ at time $t$. We compute a portfolio return $R_{p,t}$ with optional volatility scaling to $\sigma_{tgt}=15\%$, given returns $r_{i,t}$, annualised ex-ante volatility $\sigma_{i,t}$ for $N_A$ assets and transaction cost coefficient $C$.
\begin{equation}
\label{eqn:portfolio_returns}
  R_{p,t+1} = \frac{\sigma_{tgt}}{N_A}\sum_{i=1}^{N_A} \frac{w_{i,t}}{\sigma_{i,t}}\cdot r_{i,t+1} - C \left |\frac{w_{i,t}}{\sigma_{i,t}} - \frac{w_{i,t-1}}{\sigma_{i,t-1}} \right |
\end{equation}
This methodology is consistent with \cite{poh2021building,lim2019enhancing,wood2021trading,wood2022slow}. When comparing different strategies, we apply a second layer of volatility scaling to portfolio returns, so that each strategy takes the same overall risk.

As shown in \cite{lim2019enhancing,wood2021trading,wood2022slow}, the performance of LSTM and TFT Position Sizers varies across random initialisations. Therefore, to ensure robust results, we ensemble DIN models over 5 random seeds. Each model is trained with the same hyperparameters and we take a simple average of their outputs.

\section{Deep Inception Networks}
As shown in Exhibit \ref{fig:ModelArchitecture_HighLevel}, DIN models consist of a Feature Extractor (FE) followed by a Position Sizer (PS). FEs extract short-term features from standardised daily returns. We propose two novel FE architectures, OrigCIM and FlexCIM, which use customised Inception Modules to capture TS, CS and combined features. We also consider two benchmarks: DeepLOB \cite{zhang2019deeplob} only captures TS features and AxialLOB \cite{kisiel2022axial} uses axial attention instead of Convolutional Neural Networks (CNNs). 

Position Sizers capture longer-term variations. We consider two PS variants: Long Short-Term Memory (LSTM) \cite{lim2019enhancing}, and Temporal Fusion Transformer (TFT) \cite{wood2021trading, lim2021temporal}. The PS outputs portfolio weights that directly optimise Sharpe ratio, adjusted for transaction costs and correlation to Long-only. 
\begin{gather}
\label{eqn:loss}
    \mathrm{Loss} = -\sqrt{252} \cdot\frac{\mathrm{Mean}(R_{p,t})}{\mathrm{Std}(R_{p,t})} + K\cdot|\rho|
\end{gather}
where $R_{p,t}$ are volatility-scaled portfolio returns adjusted for transaction costs. As defined in Eq. \ref{eqn:portfolio_returns}, $R_{p,t}$ depends on transaction cost coefficient $C$. $R_{b,t}$ are portfolio returns of the Long-only benchmark and $\rho$ is Pearson correlation between $R_{p,t}$ and $R_{b,t}$. 

Increasing $C$ encourages the model to reduce portfolio turnover. We balance lower transaction costs against opportunity costs from excessively low turnover, by introducing a novel correlation cost term with coefficient $K$. This helps prevent the model from choosing a static set of weights that works on both train and validation sets, but is unlikely to generalise to the test set. $K$ and $C$ are tuneable hyperparameters that are fixed in the validation set. In this project, we set $K_{valid} = C_{valid} = 0$.

\subsection{Model Inputs and Outputs}
\label{section:Architecture_IO}
At each timestep $t$, DIN models predict a vector of position sizes $w_{i,t} \in [-1,1]$ for portfolio construction. DINs take 3 input matrices, each of size ($T$ x $N_A$), where $T$ is sequence length and $N_A$ is the number of assets in the portfolio.
\begin{enumerate}
    \item \textbf{Past returns $X_t$}: daily returns standardised by 63-day exponentially-weighted standard deviation $\frac{r_{i,t}}{\sigma{i,t}}$. The last row of $X_t$ is the most recent past returns, from $t-1$ to $t$.
    \item \textbf{Future returns $Y_t^{(1)}$}: $X_t$ shifted forwards by one day and scaled to $\sigma_{tgt}$. The last row is ``one-step-ahead" returns, from $t$ to $t+1$.
    \item \textbf{Volatility scaling $Y_t^{(2)}$}: volatility scaling factors $\frac{\sigma_{tgt}}{\sigma_{i,t}}$, corresponding to $Y_t^{(1)}$.
\end{enumerate}
$X_t$ is used to make predictions and $Y_t^{(1)}$, $Y_t^{(2)}$ are stacked into a single tensor $Y_t$ to evaluate the loss function. In this paper, we use a batchsize equivalent to sequence length $T$.

\subsection{Feature Extractors}
Feature Extractors (FEs) remove the need for handcrafted features and combine both TS and CS information. Starting with past returns from input $X_t$, FEs learn intermediate features which capture TS, CS and combined information, then summarise these into one feature series for each asset.

\vspace{1ex}
\textbf{DeepLOB \cite{zhang2019deeplob}}

The novel OrigCIM and FlexCIM architectures take inspiration from DeepLOB, so we include this as a benchmark. DeepLOB only extracts TS features, since it was developed for single-asset tasks. Therefore, by comparing performance against other FE variants, we can assess whether it is beneficial to incorporate CS and combined features. 

\vspace{1ex}
\textbf{AxialLOB \cite{kisiel2022axial}}

AxialLOB uses axial attention to capture long-range dependencies. This differs from DeepLOB, OrigCIM and FlexCIM, which use CNNs to capture local interactions. Although AxialLOB outperforms DeepLOB in predicting stock price movements from Limit Order Book (LOB) data, this may not be the case for DIN models. In particular, the PS can already capture attention patterns if using TFT.

\vspace{1ex}
\textbf{Original Custom Inception Module (OrigCIM)}

\begin{figure*}[!htb]
\centering
    \includegraphics[width=0.99\textwidth]{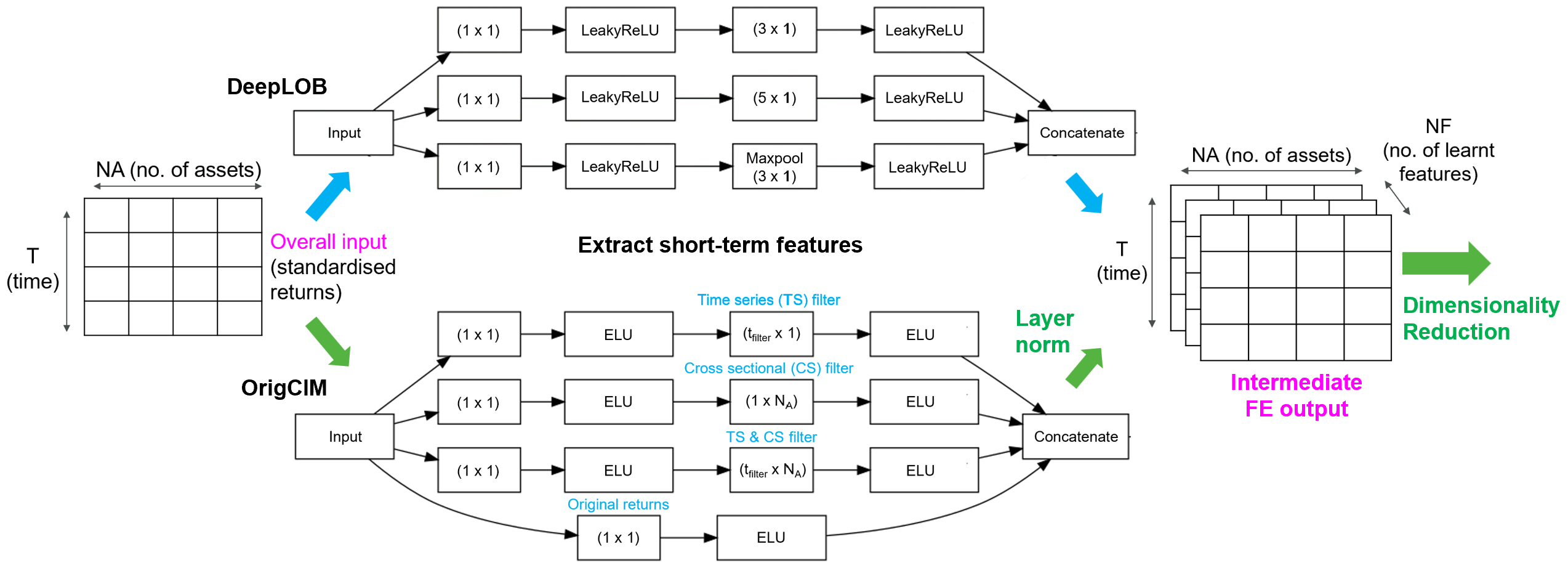}
    \caption{Original Custom Inception Module (OrigCIM) and DeepLOB Feature Extractors (FEs) for DIN models. The full FE requires dimensionality reduction, as shown in Exhibit \ref{fig:ModelArchitecture_DimRed}.}
    \label{fig:ModelArchitecture_CIM}
\end{figure*}

\begin{figure*}[!htb]
\centering
    \includegraphics[width=0.99\textwidth]{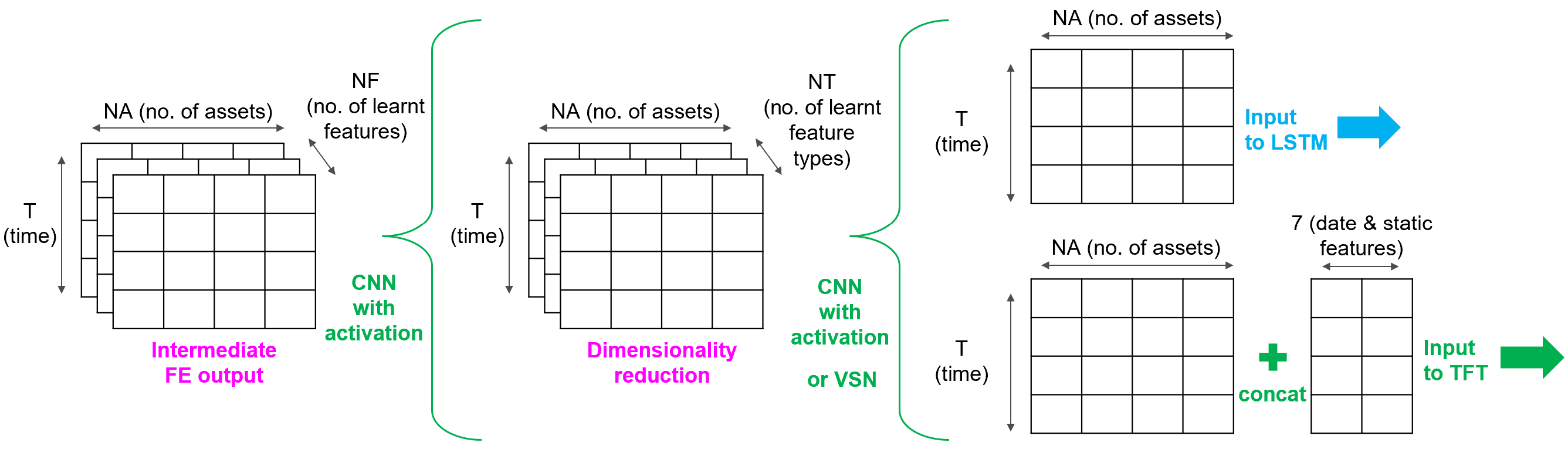}
    \caption{Dimensionality reduction for intermediate FE outputs.}
    \label{fig:ModelArchitecture_DimRed}
\end{figure*}

As illustrated in Exhibit \ref{fig:ModelArchitecture_CIM}, OrigCIM uses a single layer of filters. For CS, TS and combined features, we use filters of size ($1 \times N_A$), ($ts\_filter\_length \times 1$) and ($ts\_filter\_length \times N_A$), respectively, to capture corresponding interactions. We also use ($1 \times 1$) filters to propagate the original daily returns data to the next stage, with a learnt scaling factor. All features are learnt in parallel, using an IM structure. For each type of feature, we use $n\_filters$ filters to learn a variety of CS, TS and combined features. This intermediate feature tensor has size ($T \times N_A \times N_F$), where $N_F = N_T \times n\_filters$ and $N_T = 4$.  Finally, we use two stages of 1x1 convolutions to reduce feature dimensionality, as shown in Exhibit \ref{fig:ModelArchitecture_DimRed}.

\begin{figure*}[!htb]
\centering
    \includegraphics[width=0.99\textwidth]{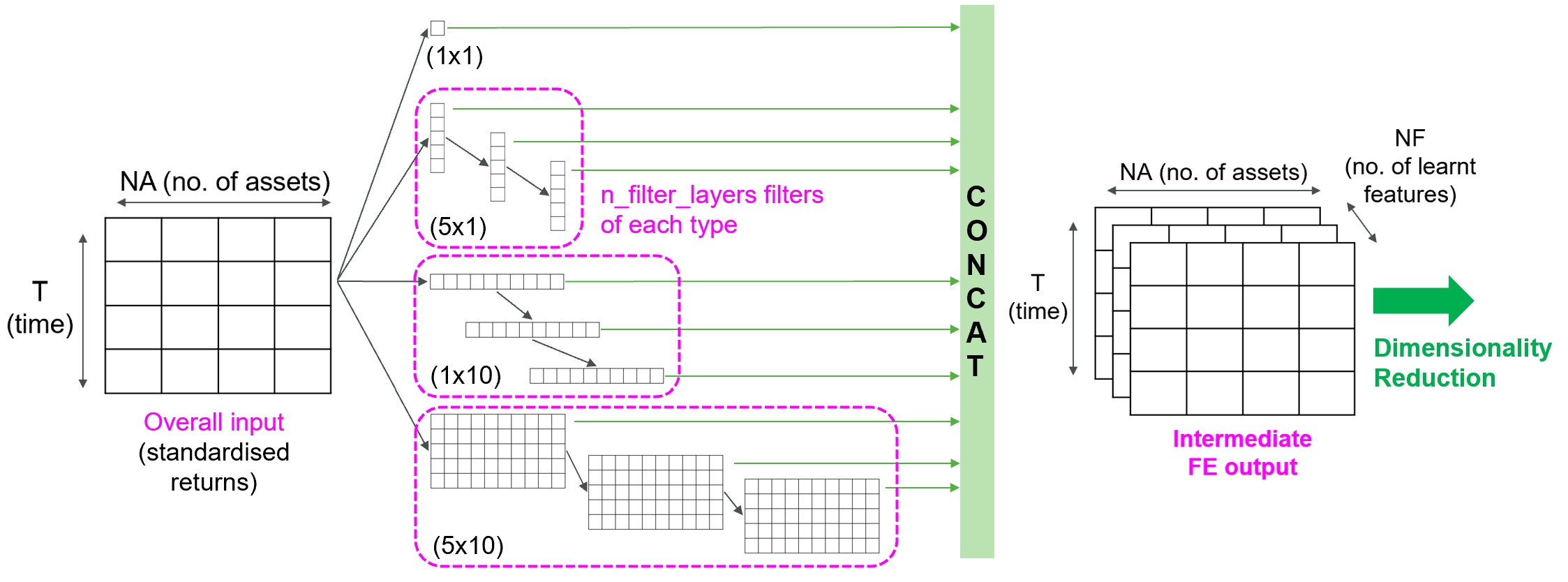}
    \caption{Flexible Custom Inception Module (FlexCIM) Feature Extractor for DIN models. ELU activations are applied after each filter. The full FE requires dimensionality reduction, as shown in Exhibit \ref{fig:ModelArchitecture_DimRed}.}
    \label{fig:ModelArchitecture_FlexibleCIM}
\end{figure*}

Compared to prior architectures, such as DeepLOB, we make the following changes to capture the desired features more efficiently:
\begin{enumerate}
    \item \textbf{No pooling}: unlike image data, fine-level spatial localisation is important for this application. Max pooling horizontally across assets is undesirable since features could be calculated with data from the wrong asset. Pooling vertically across time can be useful in higher-frequency applications to capture invariant features. However, there is a trade off with temporal accuracy. Therefore, for daily trading models, we do not apply pooling in either of the horizontal or vertical dimensions.
    \item \textbf{Larger filters}: we do not use any heuristic, for example past correlation, to order assets in $X_t$. Therefore, to calculate CS and combined features, we require filters with width equal to the number of assets $N_A$. For futures and equity data, these will be significantly larger than traditional IM filters.
    \item \textbf{Hyperparameter tuning}: rather than manually specifying $ts\_filter\_length$ (and $n\_filters$), we tune these to optimise validation loss.
    \item \textbf{Activation}: the original IMs use ReLU activation \cite{glorot2011relu}. These can be susceptible to ``dying ReLU". Both Leaky ReLU and Exponential Linear Unit (ELU) avoid this problem by scaling negative inputs to a small, non-zero value. However, ELU offers the additional advantage that large negative inputs will saturate. Therefore, we use ELU activations in OrigCIM and FlexCIM.
\end{enumerate}

\vspace{1ex}
\textbf{Flexible Custom Inception Module (FlexCIM)}

FlexCIM addresses a weakness of OrigCIM: CS and combined filters have width equal to the number of assets $N_A$. However, $N_A$ can be large and OrigCIM must be retrained whenever $N_A$ changes, for example if a new asset enters the portfolio. FlexCIM instead chains $n\_filter\_layers$ layers of small CNN filters with fixed size. As shown in Exhibit \ref{fig:ModelArchitecture_FlexibleCIM}, the CS, TS and combined filters have size ($5 \times 1$), ($1 \times 10$) and ($5 \times 10$), respectively. Due to chaining, the number of feature types $N_T$ can be large. To improve interpretability, we use a Variable Selection Network (VSN) to perform the second stage of dimensionality reduction shown in Exhibit \ref{fig:ModelArchitecture_DimRed}, instead of a CNN. The main interpretability benefit is that we can observe how the importance of the $N_T$ feature types changes over time. 

\subsection{Position Sizers}
To complement short-term features learnt by the FE, a desirable criteria for the Position Sizer (PS) is an ability to learn longer-term dependencies. Two candidate models are Long Short-Term Memory (LSTM) and Temporal Fusion Transformer (TFT). Both avoid vanishing gradients, but TFT has potential advantages for regime changes and interpretability. Further details are provided in Appendix \ref{apdx:PS_architecture}.

\section{Back-testing Details}
\subsection{Data}
We evaluate the performance of DIN models on futures, equity, cryptocurrency and FX datasets. We briefly present the datasets used in this study with further details provided in Appendix \ref{apdx:data}:
\begin{enumerate}
    \item \textbf{Futures}: continuously-linked futures contracts from Pinnacle \cite{Pinnacle} over 2000-2022. We select 50 cross-asset contracts to match prior works on \cite{lim2019enhancing,wood2021trading,wood2022slow} that use LSTM and TFT for position sizing. 
    \item \textbf{Equities}: EURO STOXX 50 (EUR50) comprises 50 blue-chip European stocks from Compustat \cite{WRDS} over 2001-2022. To avoid survivorship bias, we fix the universe to be stocks that are in EUR50 during the week prior to the start of each test set.
    \item \textbf{Cryptocurrencies}: 8 cryptocurrency spot rates from CoinMarketCap (CMC) \cite{CoinMarketCap} over 2018-2023. To avoid survivorship bias, we use Wayback Machine \cite{WaybackMachine} to select the top 25 coins by market capitalisation (MCAP) on CMC as of January 2019: the start of the first test set. Stablecoins, such as Tether and USD coin, are excluded. For forwards compatibility with work on alternative data for cryptocurrencies, we filter to a final 8 coins according to availability of alternative data on BitInfoCharts (BIC) \cite{BitInfoCharts} in January 2019. 
    \item \textbf{Foreign exchange (FX)}: 19 spot rates from Federal Reserve Board (FRB) \cite{FRB} over 2000-2023. We select currencies to match \cite{poh2022transfer}, but some pairs are not available in FRB.
\end{enumerate}

\subsection{Model Training}
Following \cite{poh2021building,lim2019enhancing,wood2021trading,wood2022slow}, we train our model with an expanding window approach with a 90\%/10\% train/validation split for hyperparameter optimisation. For futures, equities and FX, we expand our training set in 5-year intervals. The shorter cryptocurrency dataset uses 1-year increments instead. 

All DIN models are implemented in \textit{Tensorflow}. We 
customise the model training pipeline so that training losses are calculated from ``one-step-ahead" predictions of the entire batch. This ensures that the loss function in Eq. \ref{eqn:loss} accounts for the correct transaction costs and requires shuffling of data to be disabled during training. Consistent with \cite{poh2021building,lim2019enhancing,wood2021trading,wood2022slow}, we use the Adam optimiser \cite{kingma2014adam}. 

We use two hyperparameter tuning methods for DIN models: Hyperband (HB) and Bayesian Optimisation (BO). We use HB for OrigCIM and AxialLOB Feature Extractors that are slower to train. FlexCIM and DeepLOB are more efficient so we use BO. In practice, bad parameter sets can be identified after relatively few epochs on these datasets and results with both tuning methods are consistent. We train all DIN models for a maximum of 250 epochs. Early stopping after 25 epochs of non-decreasing validation loss is used to avoid overfitting to training data. The final ML model is the LM benchmark. Since LM is significantly faster than DIN models to train, we use 50 Random Search (RS) trials with early stopping after 25 epochs. Full training details and hyperparameters are provided in Appendix \ref{apdx:experiment settings}.


\section{Results and Discussion}

\subsection{Performance}

\begin{table*}[!b]
\small
\centering
\caption{\centering{
    Comparison of DIN performance against benchmarks at $VOL = 15\%$. \textcolor{blue}{Blue} indicates that \textbf{not} using volatility scaling for individual assets improves portfolio performance and vice versa for \textcolor{red}{Red}. All benchmark strategies benefit from volatility scaling. Best values in each metric are \underline{underlined}.}}
\begin{tabular}{@{\extracolsep{4pt}}lll|ccc|ccc|cc}
\toprule 
 DIN variant  & Tuning & Volatility & MAR & DDEV & MDD & Sharpe & Sortino & Calmar & CORR & BRK  \\
 or benchmark & method & scaling & \% & \% & \% &  &  &  & \% &  bps \\
  \midrule
 OrigCIM-TFT & HB & Yes & 44.5 & 9.1 & 21.6 & 2.53 & 4.19 & 2.06 & 15.6 & 2.64 \\[0.5ex] 
 OrigCIM-TFT & HB & No & \textcolor{blue}{\underline{53.8}} & \textcolor{blue}{8.7} & \textcolor{blue}{13.8} & \textcolor{blue}{\underline{2.95}} & \textcolor{blue}{\underline{5.09}} & \textcolor{blue}{3.90} & \textcolor{blue}{11.0} & \textcolor{blue}{4.82} \\[0.5ex] 
 OrigCIM-TFT & BO & Yes & 48.0 & 9.2 & 18.9 & 2.69 & 4.36 & 2.54 & 12.8 & 2.91  \\[0.5ex] 
 OrigCIM-TFT & BO & No & \textcolor{blue}{49.8} & \textcolor{blue}{\underline{8.6}} & \textcolor{blue}{\underline{10.6}} & \textcolor{blue}{2.77} & \textcolor{blue}{4.85} & \textcolor{blue}{\underline{4.70}} & \textcolor{blue}{4.7} &  \textcolor{blue}{\underline{4.83}} \\[0.5ex] 
\midrule
 OrigCIM-LSTM & HB & Yes & 28.0 & 9.2 & 23.2 & 1.72 & 2.81 & 1.21 & 13.7 & 2.56 \\[0.5ex] 
 OrigCIM-LSTM & HB & No & \textcolor{blue}{30.4} & \textcolor{blue}{9.1} & \textcolor{red}{33.8} & \textcolor{blue}{1.85} & \textcolor{blue}{3.06} & \textcolor{red}{0.90} & \textcolor{blue}{\underline{3.8}} & \textcolor{blue}{4.25} \\[0.5ex] 
\midrule
JT LS & - & Yes & 9.2 & 10.7 & 25.9 & 0.66 & 0.93 & 0.36 & \underline{-2.7} & 5.81 \\[0.5ex]
LM LS & RS & Yes & 15.6 & 10.3 & 35.8 & 1.04 & 1.52 & 0.44 & 2.8 & 1.58 \\[0.5ex]
MOP LO & - & Yes & 9.6 & 10.8 & 25.0 & 0.69 & 0.95 & 0.39 & 72.3 & 5.65 \\[0.5ex]
BAZ LO & - & Yes & 8.5 & 10.9 & 32.2 & 0.62 & 0.86 & 0.27 & 73.9 & \underline{6.97} \\[0.5ex]
CMB EW & - & Yes & 12.4 & 10.8 & 20.7 & 0.86 & 1.19 & 0.60 & 66.0 & 5.23 \\[0.5ex]
CMB VS & - & Yes & 16.6 & 10.6 & 22.3 & 1.10 & 1.55 & 0.74 & 50.0 & 3.73 \\[0.5ex]
 \midrule
 Long-only & - & Yes & 4.4 & 10.9 & 41.6 & 0.37 & 0.50 & 0.11 & - & - \\[0.5ex] 
 \bottomrule
\end{tabular}
\label{table:Results_Pinnacle_NoCosts}
\end{table*}

We evaluate strategies according to a range of performance metrics, explained in Appendix \ref{apdx:performance_metrics}. We use the abbreviations: mean annual returns ($MAR$), annualised volatility ($VOL$), downside deviation ($DDEV$), maximum drawdown ($MDD$), correlation ($CORR$), and breakeven transaction cost ($BRK$). For comparability, we scale all strategies to a target volatility of $\sigma_{tgt} = 15\%$ at the portfolio level. 

We record results for the futures dataset in Exhibit \ref{table:Results_Pinnacle_NoCosts}. The optimal DIN variant for this dataset is OrigCIM-TFT with $Sharpe$ = \textbf{2.95} and $BRK$ = \textbf{4.82 bps}. The choice of hyperparameter tuning method does not significantly affect results, since bad parameter sets can be identified after few epochs and good parameter sets are not unique. 

\begin{figure*}[h!]
\centering
  \begin{subfigure}{0.9\textwidth}
    \includegraphics[width=\textwidth]{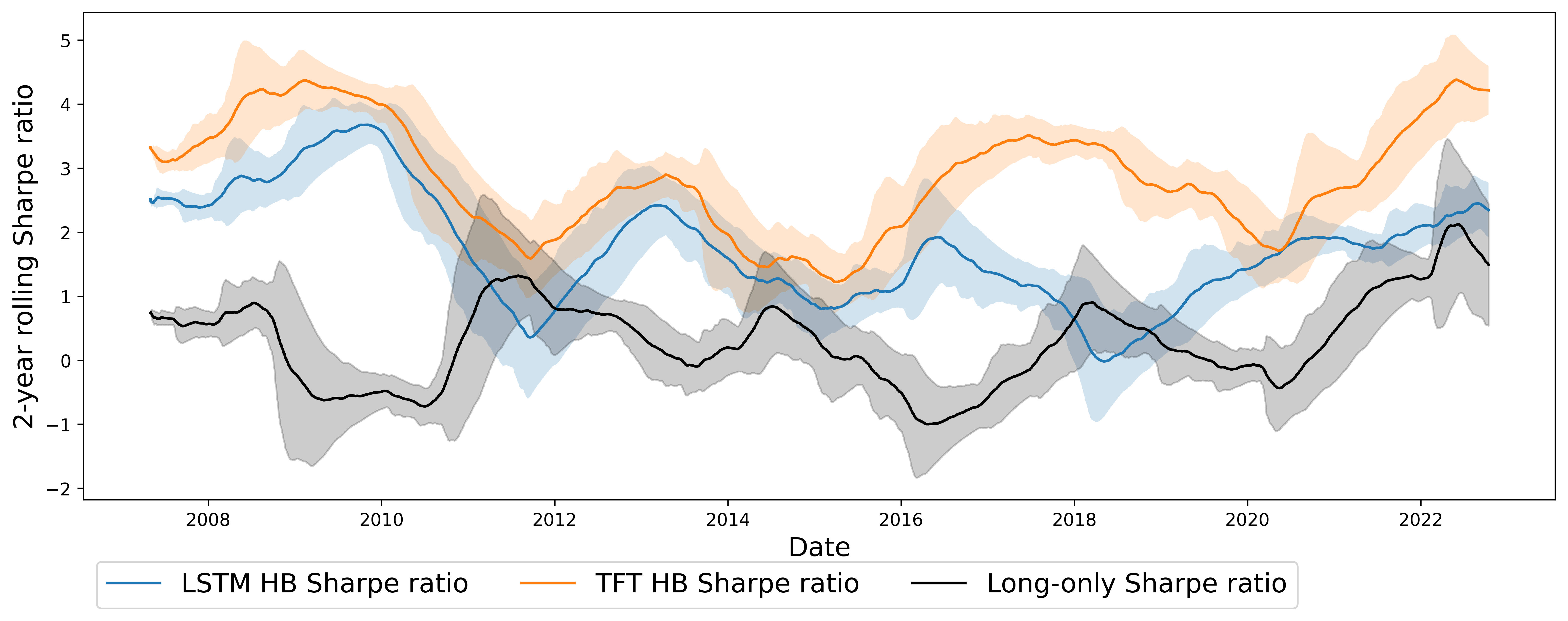}
    \caption{2-year rolling Sharpe ratio.}
    \label{fig:Results_Pinnacle_Rolling_Sharpe}
  \end{subfigure}
  
  \begin{subfigure}{0.9\textwidth}
    \includegraphics[width=\textwidth]{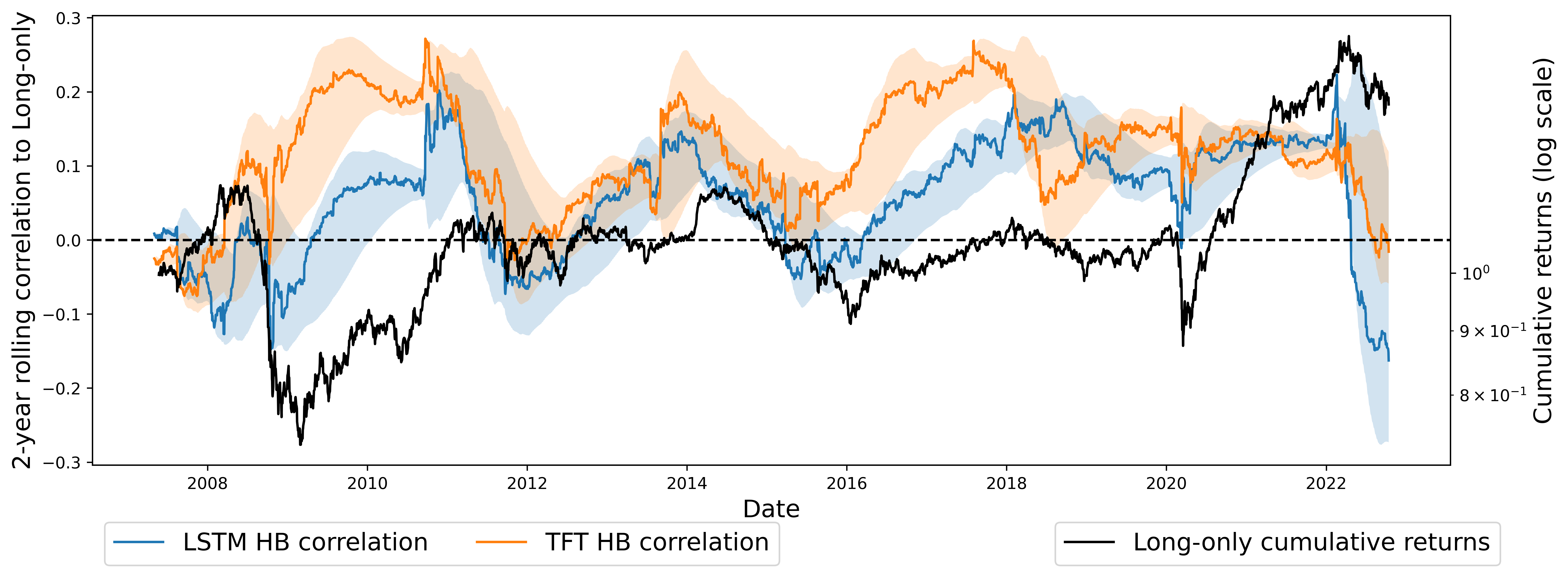}
    \caption{2-year rolling correlation.}
    \label{fig:Results_Pinnacle_Rolling_Corr}
  \end{subfigure}
  
  \caption{Rolling 2-year Sharpe ratio and correlation for DIN models compared to Long-only benchmark. The mean and standard deviation of the Sharpe ratio and correlation are computed in a rolling 63-day exponentially-weighted window. These are used to create the shaded 95\% confidence regions.} 
\label{fig:Results_Pinnacle_Rolling}
\end{figure*}

We also compare performance with and without volatility scaling at the asset level. For Long-only and other benchmarks, we find that performance without volatility scaling is strictly worse. This agrees with findings from \cite{harvey2018impactofvolscaling} that volatility scaling improves Sharpe ratio and reduces the likelihood of tail events for portfolios with exposure to risk-on assets, such as equities (Pinnacle includes equity futures). 

For DIN models, the reverse is true, and suggests that DIN models can learn a smarter weighting without explicit volatility scaling. DIN models without volatility scaling have lower hitrate $HR$, but higher profit-to-loss ratio $PNL$ and lower $DDEV$: $HR$, $PNL$ and statistical significance metrics for the Sharpe ratio are recorded in Appendix \ref{apdx:additional_results} instead of Exhibit \ref{table:Results_Pinnacle_NoCosts} due to space constraints. Therefore, the original DIN portfolio weights improve upside gains at the cost of an increased number of smaller losses. This matches the higher $BRK$, suggesting that DIN models inherently choose investment opportunities that are likely to yield returns exceeding transaction costs as a result of the turnover-opportunity cost tradeoff in the loss function. 

\begin{table*}[!htbp]
 \small
\centering
\caption{\centering{
    Strategy performance at different transaction cost coefficients $C$. \textcolor{blue}{Blue} indicates outperformance against Long-only and vice versa for \textcolor{red}{Red}. The best strategy at each $C$ is \underline{underlined}.}}
\begin{tabular}{@{\extracolsep{4pt}}llllcccccc}
\toprule   
{Strategy} & {Variant} & Tuning & {Volatility} & \multicolumn{6}{c}{Sharpe after $C$ (bps) of costs } \\
 \cmidrule{5-10} 
  & & method & scaling & 0 & 1 & 2 & 3 & 4 & 5\\
\midrule
 DIN & OrigCIM-TFT & HB & No & \textcolor{blue}{\underline{2.95}} & \textcolor{blue}{\underline{2.34}} & \textcolor{blue}{\underline{1.72}} & \textcolor{blue}{\underline{1.11}}  & \textcolor{blue}{\underline{0.50}} & \textcolor{red}{-0.11}\\
 DIN & OrigCIM-LSTM & HB & No & \textcolor{blue}{1.85} & \textcolor{blue}{1.41} & \textcolor{blue}{0.98} & \textcolor{blue}{0.54} & \textcolor{red}{0.11} & \textcolor{red}{-0.33}\\
\midrule
 JT & LS NORM\_RET\_252D & - & Yes & \textcolor{blue}{0.66} & \textcolor{blue}{0.55} & \textcolor{blue}{0.43} & \textcolor{red}{0.32} & \textcolor{red}{0.21} & \textcolor{red}{0.09} \\
 LM & LS & RS & Yes & \textcolor{blue}{1.04} & \textcolor{blue}{0.38} & \textcolor{red}{-0.28} & \textcolor{red}{-0.94} & \textcolor{red}{-1.60} & \textcolor{red}{-2.26} \\
 MOP & LO NORM\_RET\_63D & - & Yes & \textcolor{blue}{0.69} & \textcolor{blue}{0.57} & \textcolor{blue}{0.44} & \textcolor{red}{0.32} & \textcolor{red}{0.20} & \textcolor{red}{0.08} \\
 BAZ & LO MACD\_16\_48 & - & Yes & \textcolor{blue}{0.62} & \textcolor{blue}{0.53} & \textcolor{blue}{0.44} & \textcolor{blue}{0.35} & \textcolor{red}{0.26} & \textcolor{red}{0.17} \\
CMB & EW & - & Yes & \textcolor{blue}{0.86} & \textcolor{blue}{0.69} & \textcolor{blue}{0.53} & 0.37 & \textcolor{red}{0.20} & \textcolor{red}{0.03} \\
 CMB & VS & - & Yes & \textcolor{blue}{1.10} & \textcolor{blue}{0.81} & \textcolor{blue}{0.51} & \textcolor{red}{0.21} & \textcolor{red}{-0.08} & \textcolor{red}{-0.08} \\
 \midrule
 Long-only & - & - & Yes & 0.37 & 0.37 & 0.37 & 0.37 & 0.37 & \underline{0.37} \\
\bottomrule
\end{tabular}
\label{table:Results_Pinnacle_TransactionCosts}
\end{table*}

For comparison, we also record results using the LSTM Position Sizer. Although LSTM is a simpler solution, we find that performance improvements with TFT justify the added complexity. TFT-variants outperform LSTM-variants in all metrics, except correlation $CORR$. TFT-variants are able to learn portfolio weights that minimise $MDD$ without further volatility scaling, whereas this is not true for LSTM-variants. This agrees with one of the advantages of TFT-based models: long-term information can be retained further and across regime changes, which may help TFT-variants avoid larger losses, whilst LSTM-variants require additional help. Exhibit \ref{fig:Results_Pinnacle_Rolling} shows rolling 2-year Sharpe ratio and correlations against Long-only benchmark without volatility scaling. The TFT-variant outperforms both the LSTM-variant and Long-only benchmark in all time periods. The correlation to Long-only for both DIN variants also remains stably low between $\pm 30\%$.

A key consideration for daily trading strategies is costs. Exhibit \ref{table:Results_Pinnacle_TransactionCosts} shows Sharpe ratios for ensembled DIN models and benchmark strategies for a range of transaction cost coefficients $C$. We assume 0 transaction costs for the Long-only benchmark. All benchmarks are shown with volatility scaling, as this improves $Sharpe$. DIN models, specifically TFT-based variants with HB and BO tuning, outperform all benchmarks up to $C$ = \textbf{4 bps}. At greater transaction costs, Long-only has highest $Sharpe$. 

\subsection{Interpretability}

Interpretability of investment decisions is critical for meeting regulatory requirements, maintaining investor confidence and diversifying risk from similar factors. The ability to understand how features contribute to model predictions can also be used to debug or improve models. In the context of systematic trading strategies, we require interpretability methods that can provide local explanations for individual predictions (trading decisions), and expect the importance of features to vary over time.

Interpretability techniques can be split into two broad categories: model-agnostic and model-specific. Model-agnostic methods have the advantage that they can be applied to any ML model. A popular method that supports local explanations is Local Interpretable Model-agnostic Explanations (LIME) \cite{ribeiro2016LIME}. LIME fits a simple model, such as linear regression, to predictions generated by the original, more complex model in a local region of the input space. Shapley Additive Explanations (SHAP) \cite{lundberg2017SHAP} is sometimes preferred over LIME due to its consistency and local accuracy properties. SHAP uses a game theoretic approach to calculate feature contribution by comparing predictions made using different combinations of features. 

However, neither method is suitable for DINs. Both LIME and SHAP perform best if data distributions remain reasonably constant over time. This may be sensible in tasks such as object recognition, but financial data is notoriously non-stationary. For example, feature importance may differ between risk-on and risk-off regimes, or trending and reverting regimes. Another problem is the high dimensionality of the DIN input. The input matrix $X_t$ has size ($T \times N_A$ = $100 \times 50$ = 5000) ``pixels" for the futures portfolio. If we treat each ``pixel" as a feature, it is difficult to build intuition of what these features mean, and to differentiate noise. 

Instead, we use model-specific methods. Due to the large number of trainable parameters in DIN models, it is infeasible to visualise all of the neural network weights. However, there are two subsets of trained parameters, which are particularly insightful: attention and VSN weights.

\vspace{1ex}
\textbf{Attention}

\begin{figure*}[!b]
  \centering
    \includegraphics[width=0.98\textwidth]{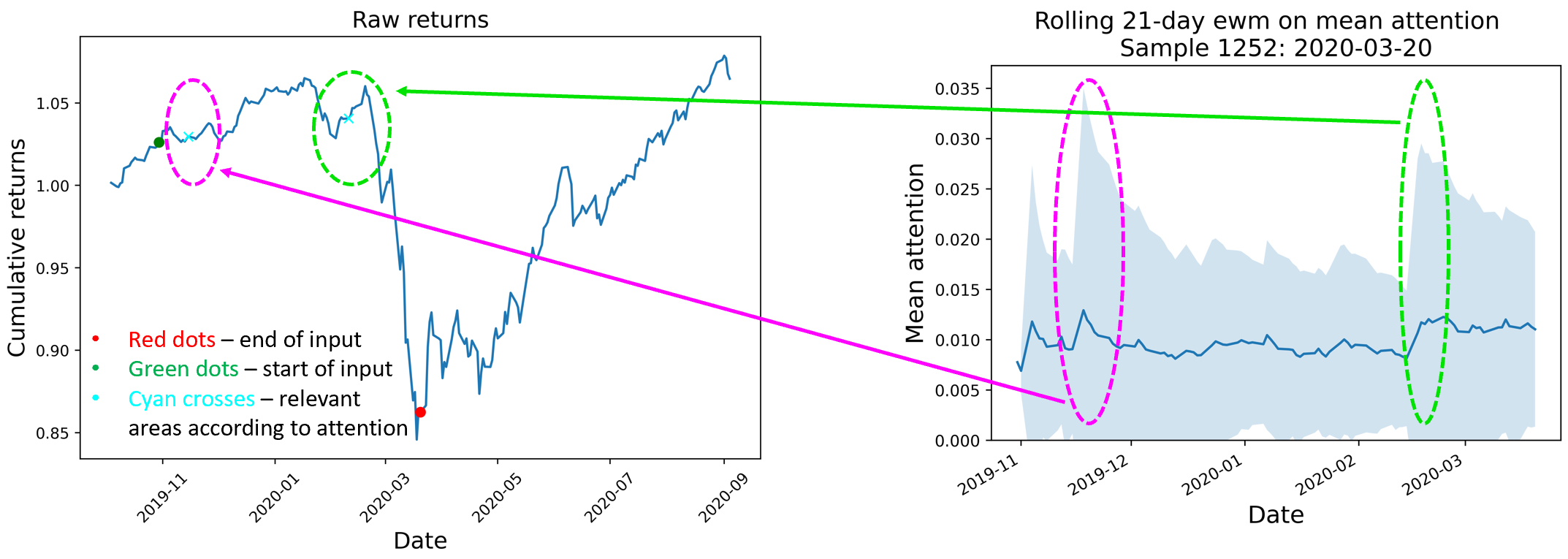}
    \caption{Attention patterns for a DIN variant (FlexCIM-TFT) on Pinnacle data. Peaks in attention correspond to locations in the long-only cumulative returns with similar shape to the prediction timestep.}
    \label{fig:Attention_Example}
\end{figure*}

\begin{figure*}[!htbp]
  \centering
  \begin{subfigure}{0.49\textwidth}
    \includegraphics[width=\textwidth]{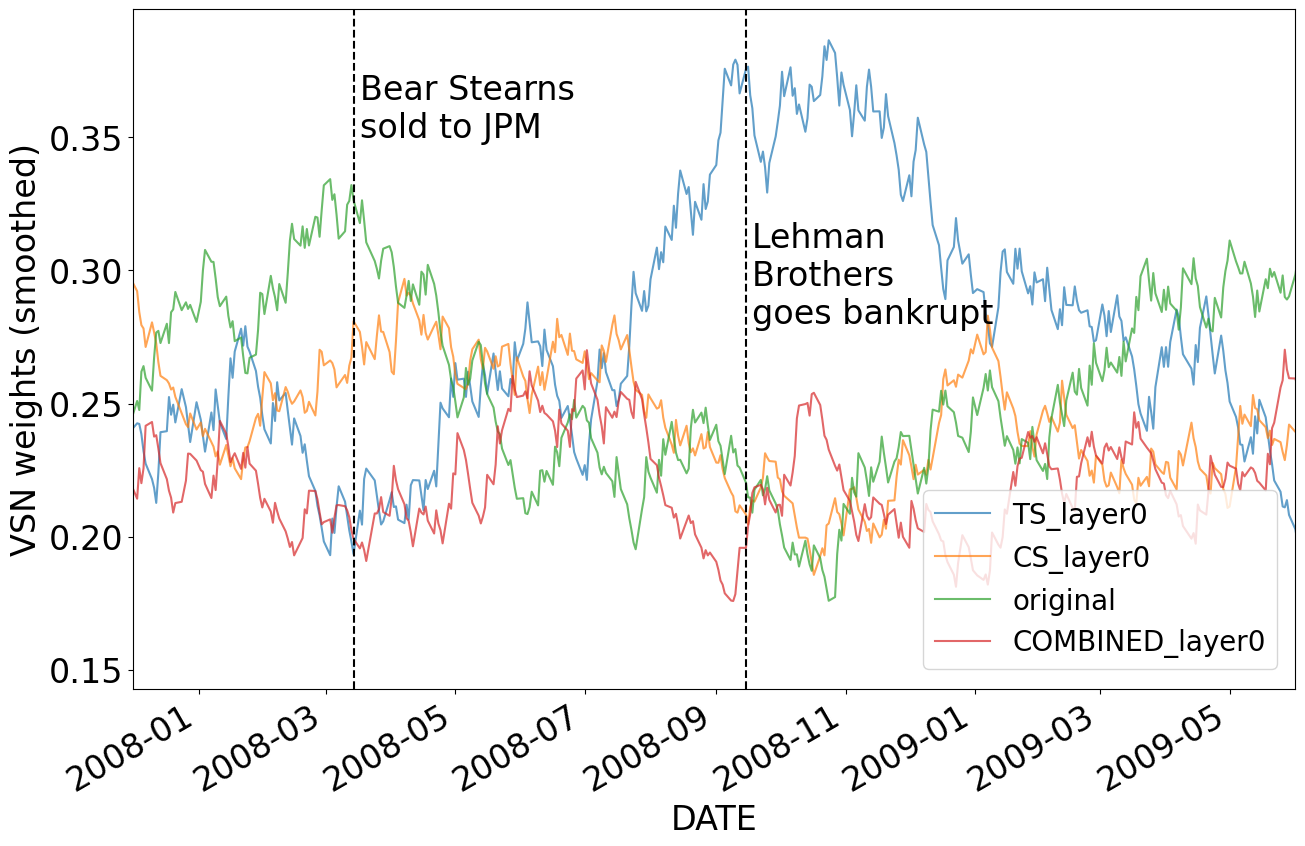}
    \caption{Great Financial Crisis (GFC).}
    \label{fig:VSN_Example1}
  \end{subfigure}
  \hspace{0.01\textwidth}%
  \begin{subfigure}{0.49\textwidth}
    \includegraphics[width=\textwidth]{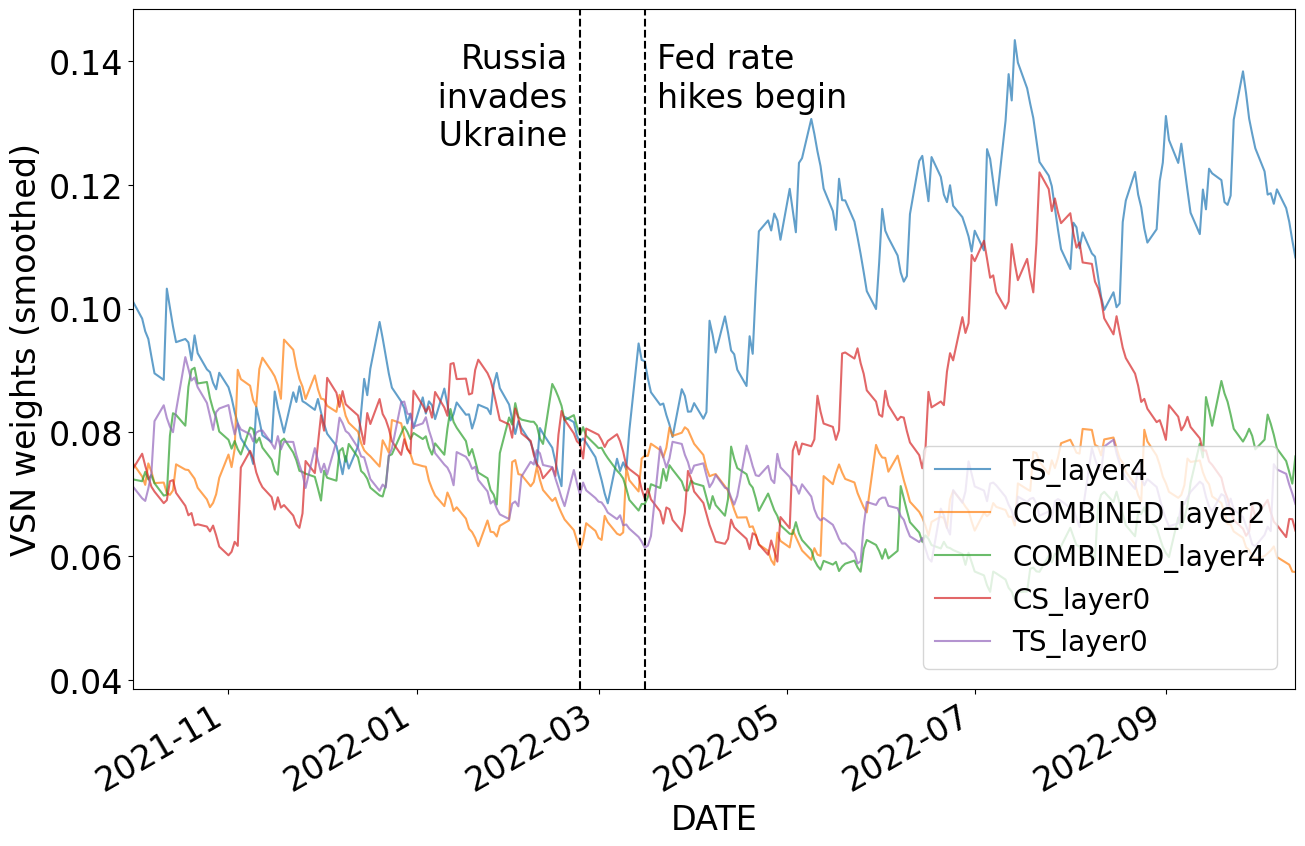}
    \caption{Russian invasion of Ukraine.}
    \label{fig:VSN_Example2}
  \end{subfigure}
  
  \caption{VSN weights from FlexCIM, smoothed with a 63-day rolling exponentially-weighted window.} 
\label{fig:VSN_Example}
\end{figure*}

We consider mean attention $\Tilde{A}$ from the TFT Position Sizer. For visualisation, we smooth with a 63-day rolling exponentially-weighted window and plot a 95\% confidence interval. In general, attention is higher at previous times in the input sequence where the market (Long-only cumulative returns) follows a similar pattern to the prediction timestep. A specific example is illustrated in Exhibit \ref{fig:Attention_Example}. Following the March-2020 market crash due to COVID-19, attention peaks correspond to prior periods where there has been a post-dip recovery, rather than the bear market directly preceding the prediction timestep.

\vspace{1ex}
\textbf{Variable Selection Networks (VSNs)}

VSN weights $\nu_{t,j}$ from the FlexCIM FE can be used to visualise how the importance of different types of extracted features $j$ varies over time $t$. There is another VSN inside the TFT PS, which is available regardless of chosen FE. However, the VSN inside TFT is less interpretable, since it operates on condensed features (following dimensionality reduction), which do not have obvious spatial-temporal semantics.

As a case study, we consider the periods surrounding key market events: the Great Financial Crisis (GFC) of 2007-2009, as well as the Russia-Ukraine war and subsequent initiation of historically aggressive tightening by the Federal Reserve (Fed). In Exhibit \ref{fig:VSN_Example}, we plot $\nu_{t,j}$, smoothed with exponentially-weighted windows for visualisation. We also isolate the top 5 features by mean importance to aid interpretability. 

In both examples, we find a clear change in feature importance after key events. During GFC in Exhibit \ref{fig:VSN_Example1}, there is rotation away from short-term (1-day) ``original" features towards longer (5-day) TS features immediately after the Bear Stearns takeover, peaking after the Lehman Brothers bankruptcy. After the 2022 Fed rate hikes shown in Exhibit \ref{fig:VSN_Example2}, the longest TS feature (with receptive field of 21 days) quickly dominates. This may be due to increased trending behaviour as investors rotate away from risk assets.

\subsection{Results on Other Asset Classes}

\begin{table*}[!hbp]
\small
\centering
\caption{\centering{ 
        Comparison of DIN model $Sharpe$ and $MDD$ with different Feature Extractors (FEs) on futures, equity, cryptocurrency, FX datasets. For comparability, all models use TFT Position Sizer (PS) and are scaled at portfolio level to $VOL = 15\%$. Best values per asset class are \underline{underlined}.}}
\begin{tabular}{@{\extracolsep{4pt}}lll|cccc|cccc}
\toprule  
 DIN variant & Tuning & Volatility & \multicolumn{4}{c}{ $Sharpe$ } & \multicolumn{4}{c}{ Maximum Drawdown, $MDD$ $(\%)$ }   \\
  & method & scaling & Futures & Equity & Crypto & FX & Futures & Equity & Crypto & FX \\
 \midrule
 OrigCIM-TFT & HB & No & \underline{2.95} & 1.15 & 0.06 & 1.27 & 13.8 & \underline{23.9} & 26.4 & 26.4 \\[0.5ex]  
 FlexCIM-TFT & BO & No & 2.33 & 0.92 & 0.25 & \underline{1.49} & \underline{13.5} & 27.9 & 32.5 & 27.2 \\[0.5ex]
 DeepLOB-TFT & BO & No & 2.32 & \underline{1.20} & \underline{1.02} & 1.19 & 20.1 & 24.4 & \underline{14.4} & 23.0 \\[0.5ex] 
 AxialLOB-TFT & HB & No & 0.36 & 0.7 & -0.10 & 1.17 & 38.6 & 26.1 & 37.2 & \underline{22.1}  \\[0.5ex] 
\midrule
 Long-only & - & No & 0.29 & 0.34 & 0.72 & 0.42 & 54.2 & 40.2 & 25.4 & 45.3 \\[0.5ex] 
 \bottomrule 
\end{tabular}
\label{table:Results_AllDatasets_FE_SharpeMDD}
\end{table*}

\begin{table*}[!hbp]
\small
\centering
\caption{\centering{ 
        Comparison of DIN model $BRK$ and $CORR$ with different Feature Extractors (FEs) on futures, equity, cryptocurrency, FX datasets. For comparability, all models use TFT Position Sizer (PS) and are scaled at portfolio level to $VOL = 15\%$. Best values per asset class are \underline{underlined}.}}
\begin{tabular}{@{\extracolsep{4pt}}lll|cccc|cccc}
\toprule  
 DIN variant & Tuning & Volatility & \multicolumn{4}{c}{ Breakeven cost, $BRK$ $(bps)$ } & \multicolumn{4}{c}{ Correlation to Long-only, $CORR$ $(\%)$ }   \\
  & method & scaling & Futures & Equity & Crypto & FX & Futures & Equity & Crypto & FX \\
 \midrule
OrigCIM-TFT & HB & No & \underline{4.82} & 2.36 & 0.93 & 1.62 & 11.0 & \underline{9.7} & 5.0 & -7.4 \\[0.5ex]  
 FlexCIM-TFT & BO & No & 4.31 & 1.99 & 5.06 & 1.56 & 11.8 & 11.3 & -4.2 & -6.9 \\[0.5ex]
 DeepLOB-TFT & BO & No & 4.05 & \underline{2.93} & \underline{29.91} & 2.00 & 7.6 & 10.4 & 9.5 & \underline{2.2} \\[0.5ex] 
 AxialLOB-TFT & HB & No & 0.71 & 1.30 & -2.41 & \underline{2.76} & \underline{-3.0} & 11.0 & \underline{2.8} & 24.7  \\[0.5ex] 
\bottomrule  
\end{tabular}
\label{table:Results_AllDatasets_FE_BRKCORR}
\end{table*}

We compare DIN performance with different FEs on various asset classes in Exhibits \ref{table:Results_AllDatasets_FE_SharpeMDD} and \ref{table:Results_AllDatasets_FE_BRKCORR}, presenting key metrics of $Sharpe$, $MDD$, $BRK$, $CORR$. Appendix \ref{apdx:additional_results} contains metrics for statistical significance.
Based on $Sharpe$ before transaction costs, FEs that incorporate CS information (OrigCIM, FlexCIM) outperform on futures and FX. DeepLOB (only TS features) performs best on equities and cryptocurrencies. This may be because there are less useful CS patterns in equities and cryptocurrencies, due to positive correlation between all assets.

In particular, only DeepLOB FE outperforms the Long-only benchmark for cryptocurrencies. This could be due to the limited data for cryptocurrencies in comparison to other assets. In contrast to profitable DIN models on other asset classes, DeepLOB FE on cryptocurrencies produces a low-$Sharpe$ strategy (only \textbf{1.4x} increase over Long-only $Sharpe$), with high robustness to transaction costs at $BRK$ = \textbf{29.91 bps}. Despite lower turnover, $CORR$ is low at \textbf{9.5\%} and $MDD$ is significantly improved at \textbf{14.4\%}.

\subsection{Computational Complexity}

\begin{figure*}[!htbp]
\centering
  \begin{subfigure}{0.47\textwidth}
    \includegraphics[width=\textwidth]{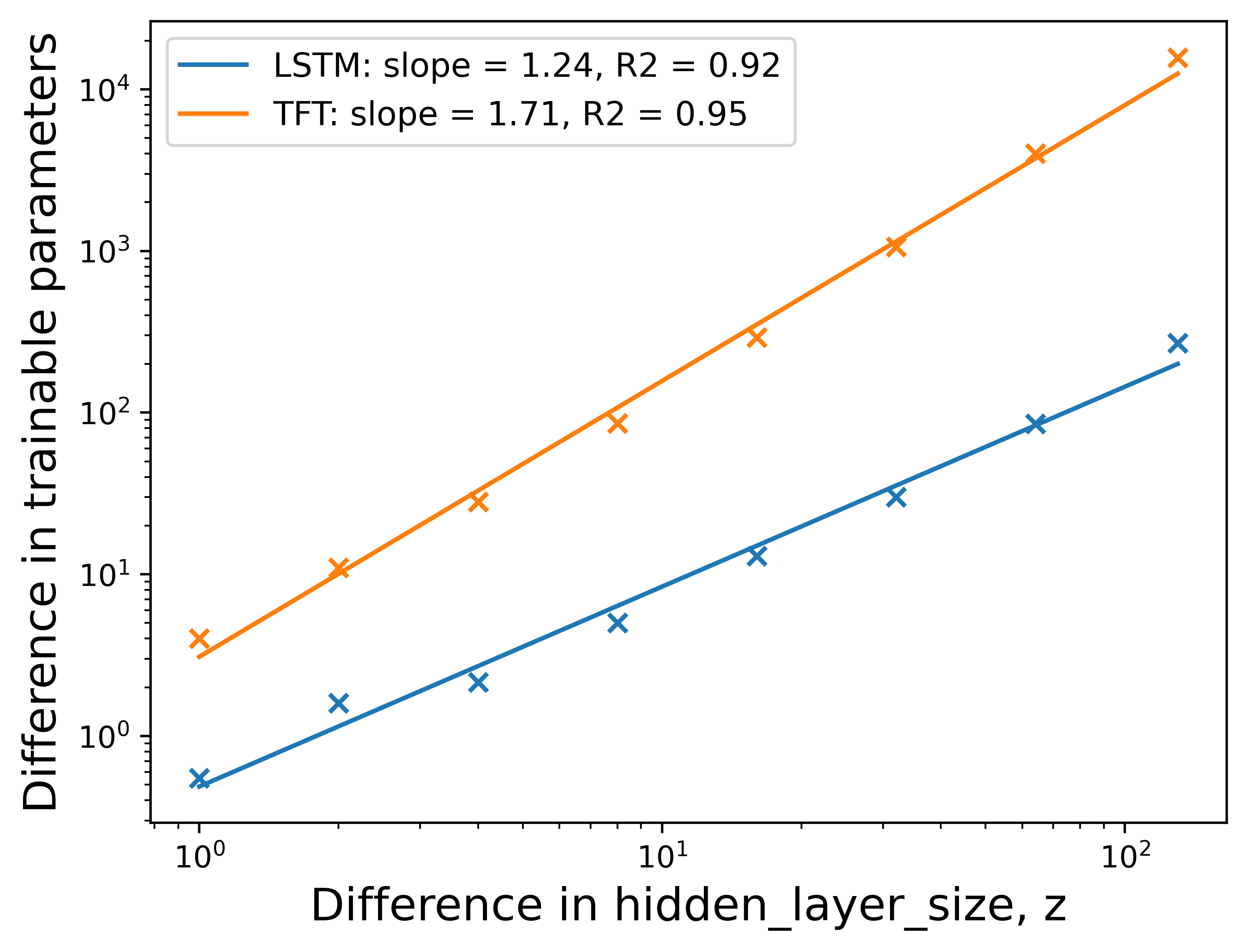}
    \caption{Complexity of PS variants.}
    \label{fig:Complexity_PS}
  \end{subfigure}
  \hspace{0.04\textwidth}%
  \begin{subfigure}{0.47\textwidth}
    \includegraphics[width=\textwidth]{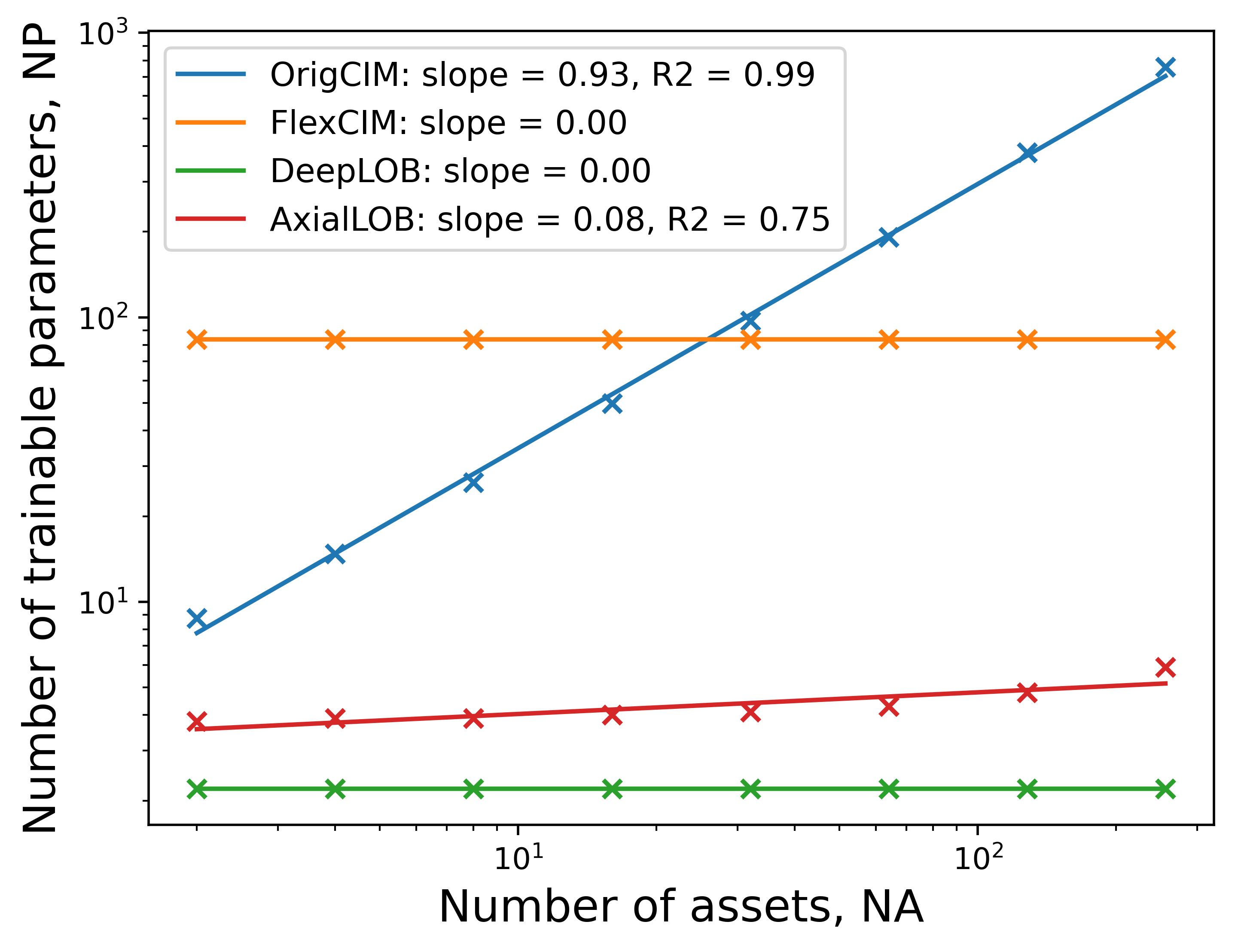}
    \caption{Complexity of FE variants.}
    \label{fig:Complexity_FE}
  \end{subfigure}
  
  \caption{DIN model complexity, according to trainable parameters. To assess PS complexity, we construct the complete DIN model and remove the effect of the FE component through differencing. To assess FE complexity, we construct each FE as a standalone model without a PS component.} 
\label{fig:DIN_Complexity}
\end{figure*}

For the DIN Position Sizer (PS) component, the number of trainable parameters $N_P$ increases as a function of $N_H = hidden\_layer\_size$. To remove the effect of the Feature Extractor (FE), we use differencing: this considers the average increase in $N_P$ for a difference in $N_H$. Plotting these differences on a log-log scale in Exhibit \ref{fig:Complexity_PS}, the complexity of LSTM and TFT PS-variants are $\mathcal{O}(z^{1.24})$ and $\mathcal{O}(z^{1.71})$, with respect to $z = \Delta N_H$.

To assess Feature Extractor (FE) complexity, we consider how the number of trainable parameters $N_P$ in the FE component depends on $n\_filters$ and one of the following: $L_1 = ts\_filter\_length$ (OrigCIM), $L_2 = n\_filter\_layers$ (FlexCIM), $L_3 = n\_axial\_heads$ (AxialLOB). DeepLOB has one fewer hyperparameter than other FEs. In Exhibit \ref{fig:Complexity_FE}, we plot average $N_P$ for increasing $N_A$ on a log-log plot. Here, average denotes the mean of upper and lower limits of $N_P$, according to the range of allowable hyperparameter values for $L_1$, $L_2$, $L_3$ defined in Appendix \ref{apdx:experiment settings}. We fix $n\_filters=16$, since this parameter is shared between all FEs.

The key result is that the complexity of FlexCIM is independent of $N_A$, whereas OrigCIM has approximately linear dependence on $N_A$. This is despite capturing local CS and combined features, which are not extracted by DeepLOB and AxialLOB. 

\section{Conclusions}
We propose an end-to-end Deep Learning framework for systematic trading strategies. Extending prior work on single assets, Deep Inception Network (DIN) models produce investment decisions that optimise $Sharpe$ loss, adjusted for transaction costs and absolute correlation, for the entire portfolio. We demonstrate that DIN models outperform a selection of traditional and ML benchmarks at all transaction costs up to \textbf{4 bps} on cross-asset futures data from 2005-2022. Beyond this point, buy-and-hold strategies such as Long-only perform best.

A key innovation is to balance the generality of DIN models against overfit, whilst maintaining interpretability. This is achieved by the Feature Extractor (FE) module. The researcher can design and systematically compare the performance of different FEs. These capture specific types of behaviour that have semantic meaning, which helps avoid overfit. For example, time series (TS), cross sectional (CS) and combined filters in the proposed OrigCIM and FlexCIM FEs may capture momentum, spread and lead-lag effects, respectively. However, the exact features are learnt automatically from data and not hand-crafted. This also enables generalisation to different asset classes, by choosing suitable FEs. For example, DeepLOB FE, which only captures TS information, performs best on equity and cryptocurrency datasets, where almost all assets are positively correlated. In this situation, the CS features learnt by OrigCIM and FlexCIM are less useful.

DINs are also inherently interpretable. VSN weights in FlexCIM FEs provide local explanations of dynamic feature importance over time. For example, we provide case studies where long TS features gain importance during crisis events, such as the Great Financial Crisis, suggesting increased focus on trending features and less dependence on previously learnt CS relationships. This is complemented by attention weights, which highlight previous time points that are most important to the current prediction. In general, greater weight is given to previous return patterns that resemble the current environment, allowing for implicit regime conditioning. In contrast, common model-agnostic methods, such as LIME and SHAP, are difficult to interpret due to high dimensionality of the input data. 

The current DIN framework only learns features from price action data. One avenue of future work is to improve generality by extending the input dimensionality from ($T \times N_A$) to ($T \times N_A \times N_I$), for sequence length $T$, number of assets $N_A$ and number of input features $N_I$. The additional input features could include technical indicators (e.g. volume), low-frequency fundamental data for equities (e.g. book-to-market), or daily frequency alternative data for cryptocurrencies (e.g. hashrate and Google trends). A possible method is to add an input stage to blend the input features, using a VSN or CNN filters.

\section{Acknowledgements}
We would like to thank the Oxford-Man Institute of Quantitative Finance for computing support. This paper is derived from a Master's dissertation on ``Combining Cross Sectional and Time Series Systematic Trading Strategies", written by Tom Liu and supervised by Professor Zohren and Professor Roberts.

\bibliographystyle{IEEEtran}
{\footnotesize
\bibliography{references}
} 
\clearpage

\newpage
\appendix

\subsection{Benchmark Strategies}
\label{apdx:benchmarks}

Here, we describe our benchmark construction rules in more detail. In general, we choose the variant of each benchmark with highest $Sharpe$. This is a conservative overestimate of benchmark performance, due to relying on ex-post information.

\vspace{1ex}
\textbf{Long-only}

This benchmark represents market returns and assigns equal weight $w_{i,t} = 1$ to all assets $i$. At each time $t$, we use the same universe of assets as the DIN model. 

\subsubsection{Cross Sectional (CS) Benchmarks}
For CS strategies, we create a portfolio that longs/shorts the decile of assets $i$ with highest/lowest signal $s_{i,t}$ at each time $t$. We consider separate results for long-only (LO) and long-short (LS) portfolios, such that
\begin{equation}
    w_{i,t} = 
    \begin{cases}
      1, & \text{if $\mathrm{Percentile}(s_{i,t}) \geq 90$}\\
      -1, & \text{if $\mathrm{Percentile}(s_{i,t}) \leq 10$ and LS}\\
      0, & \text{otherwise}
    \end{cases} 
\end{equation}
and choose the option with highest Sharpe ratio.

\vspace{1ex}
\textbf{Jegadeesh \& Titman (JT) \cite{jegadeesh1993returns}}

JT is the seminal work on CS momentum. The authors propose ranking assets based on their returns in the past 3-12 months. In this paper, we modify the signal $s_{i,t} = r_{i,t}^{(\hat{k})}$ to be returns over past $\hat{k} \in \{5, 21, 63, 126, 252\}$ days, where $\hat{k}$ optimises the Sharpe ratio.

\vspace{1ex}
\textbf{LambdaMART (LM) \cite{poh2021building}}

Learning to Rank algorithms have been shown to improve the Sharpe ratio of CS momentum strategies. Following results from \cite{poh2021building}, we use the best-performing LambdaMART (LM) model \cite{burges2010ranknet} to rank assets. The signal $s_{i,t} = \mathrm{rank}(i,t)$ is ordered such that assets $i$ with larger rank are expected to outperform assets with smaller rank in the future. The inputs are standardised returns over \{1, 5, 21, 63, 126, 252\} days and MACD indicators. We mirror the MACD implementation in \cite{baz2015dissecting} with 3 sets of timescales $(S_k,L_k)  = \{(8,24), (16,48), (24,96)\}$, and a combined MACD indicator weighted by a response function.

\subsubsection{Time Series (TS) Benchmarks}
For TS strategies, we create a portfolio that buys/shorts assets with positive/negative signal strength. Again, we consider both long-only (LO) and long-short (LS) portfolios, choosing the option with highest Sharpe ratio.
\begin{equation}
\label{eq:TS_weights}
    w_{i,t} = 
    \begin{cases}
      \mathrm{sign}(s_{i,t}), & \text{if LS}\\
      \max(0, \mathrm{sign}(s_{i,t})), & \text{if LO}\\
    \end{cases} 
\end{equation}

\textbf{Moskowitz, Ooi \& Pedersen (MOP) \cite{moskowitz2012time}}

MOP is the original TS momentum strategy. The authors propose a score based on past 12-month returns for a strategy with monthly investment horizon. In this project, we use the same signal $s_{i,t} = r_{i,t}^{(\hat{k})}$ as the modified JT strategy, but portfolio weights are determined by the TS method.

\vspace{1ex}
\textbf{Baz et al. (BAZ) \cite{baz2015dissecting}}

BAZ proposes using MACD indicators to capture the cross-over of exponentially-weighted moving averages (EWMA) of the price $p_{i,t}$. When the EWMA with the short timescale $S_k$ is greater than the EWMA with the longer timescale $L_k$, then we expect the asset to have upwards momentum and vice versa. Therefore, the BAZ signal is $s_{i,t} = \mathrm{MACD}(\hat{k})$, where we choose the MACD indicator that maximises portfolio Sharpe ratio.

\subsubsection{Combined (CMB) Benchmarks}
\label{section:Methods_Benchmarks_CMB}
To combine TS and CS strategies, we weight returns of the JT, LM, MOP, BAZ strategies, indexed by $m$, inversely proportional to the 63-day exponentially-weighted standard deviation $\sigma_{p,t}^{m}$ of portfolio returns $R_{p,t}^{m}$. We call this the CMB Volatility Scaled (VS) strategy.
\begin{equation}
    w_{m,t} = \frac{ 1/\sigma_{p,t}^{m} }{ \sum_{m=1}^{M} 1/\sigma_{p,t}^{m} }
\end{equation}
We additionally consider a CMB Equal Weight (EW) portfolio where $w_{m,t} = 1/M = 0.25$.

\subsection{Position Sizer Architectures}
\label{apdx:PS_architecture}

\vspace{1ex}
\textbf{Long Short-Term Memory (LSTM) \cite{lim2019enhancing, hochreiter1997LSTM}}
\label{section:Architecture_LSTM}

\begin{figure}[!htb]
\centering
    \includegraphics[width=0.45\textwidth]{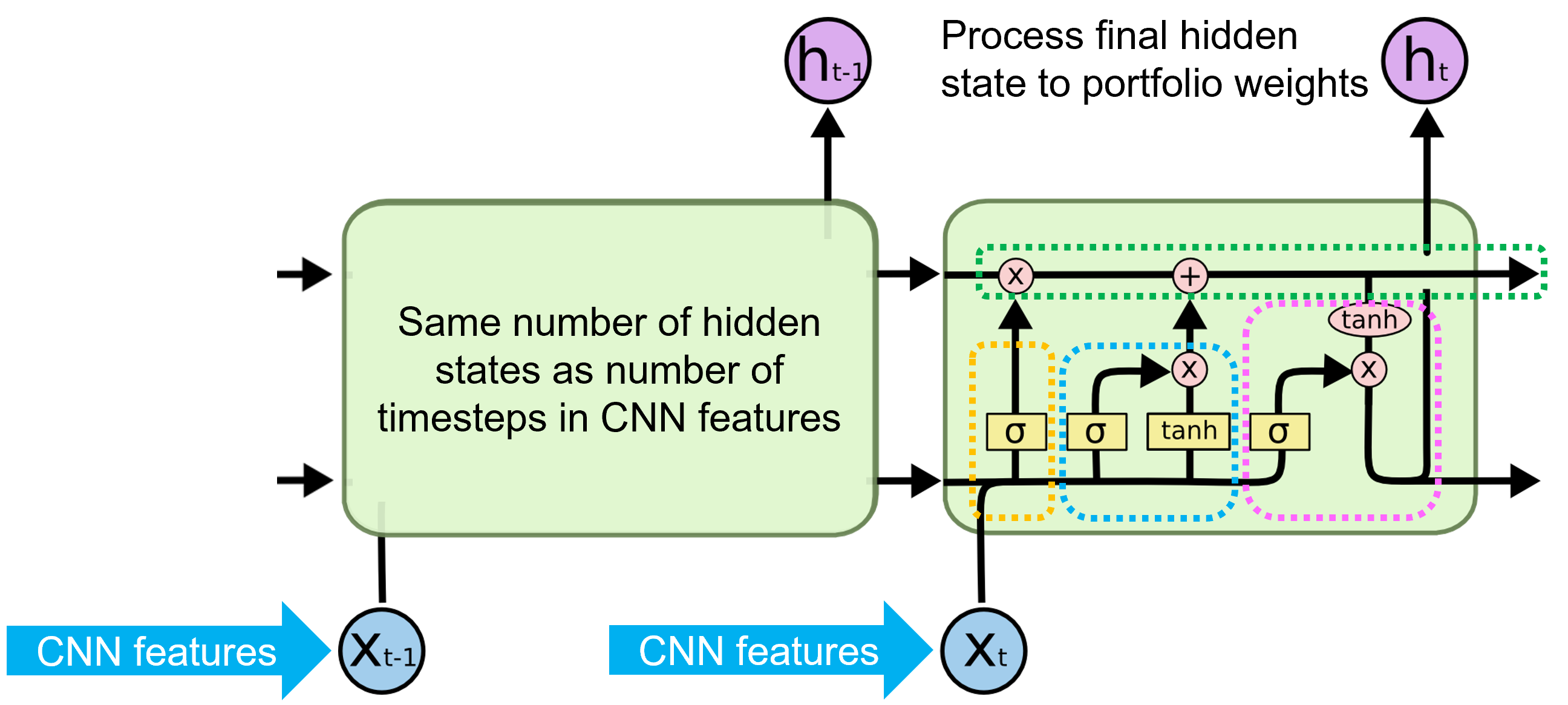}
    \caption{LSTM cell, adapted from \cite{LSTM_diagram}.}
    \label{fig:ModelArchitecture_LSTM}
\end{figure}

As shown in Exhibit \ref{fig:ModelArchitecture_LSTM}, each timestep of the FE output is the input for a corresponding LSTM cell. 

The \textcolor{teal}{cell state} allows the LSTM to preserve memory, controlled by gates: a fully connected layer with \textit{sigmoid} activation followed by pointwise multiplication. These control how much data to keep or zero out. In a standard LSTM cell, there are three gates: the \textcolor{orange}{forget gate} filters the previous cell state. The new input is processed with a fully connected layer with \textit{tanh} activation to scale outputs between $[-1,1]$, then filtered with the \textcolor{cyan}{input gate} before being added to the new cell state. The final \textcolor{magenta}{output gate} scales and filters the new cell state to create the hidden state for the timestep that corresponds to the input.

We use $N_H = hidden\_layer\_size$ LSTM cells, resulting in $N_H$ hidden states for each timestep. A fully connected layer processes the hidden units into the desired output shape of $N_A$ portfolio weights at each timestep, and the last/most recent row is used for portfolio construction. Full implementation details are provided in \cite{lim2019enhancing}.

\vspace{1ex}
\textbf{Temporal Fusion Transformer (TFT) \cite{wood2021trading, lim2021temporal}}
\label{section:Architecture_TFT}

Although the LSTM can provide a degree of long-term memory, TFT methods have potential advantages. First, LSTM memory works by selectively forgetting and replacing parts of the cell state. Therefore, it is prone to forgetting information longer in the past, especially after regime changes. TFT is an attention-LSTM hybrid model that addresses this problem: an attention layer spans all timesteps of the LSTM output and focuses on relevant information, even if it is subsequently forgotten by the LSTM. 

Another challenge for LSTM-variants of DIN models is interpretability. Traditional interpretability methods like LIME \cite{ribeiro2016LIME} and SHAP \cite{lundberg2017SHAP}, do not deal well with varying importance over time. The TFT model in particular has improved interpretability via importance weights in Variable Selection Networks and attention weights. However, we retain the LSTM-variant for comparison as this is a simpler model, with lower memory consumption and faster training.

Following \cite{wood2021trading}, we alter the original TFT model \cite{lim2021temporal} with the following changes.
\begin{enumerate}
    \item \textbf{No known future inputs}: inputs for the TFT models are past returns, date features and a static context - in this case, an integer to identify the dataset. The chosen date features are $days\_since\_start$, $month$, $year$, $day\_of\_week$, $day\_of\_month$, $week\_of\_year$. To avoid using future knowledge, date features are normalised to the [0,1] range using a fixed integer. No known future data is currently used. 
    \item \textbf{Predicts up to $t+1$}: rather than multi-horizon forecasts of the original TFT, we only predict portfolio weights up to the next timestep. Since DIN models use daily frequency data to make daily investment decisions, we only use the ``one-step-ahead" prediction in portfolio construction.
    \item \textbf{Sharpe loss function}: we replace the quantile loss function with the formulation in Eq. \ref{eqn:loss}.
\end{enumerate}

Full details of the TFT implementation are presented by \cite{wood2021trading}. Hyperparameters for all DIN and benchmark models are given in Appendix \ref{apdx:experiment settings}.

\subsection{Performance Metrics}
\label{apdx:performance_metrics}

We evaluate strategies according to a range of performance metrics, split into the following categories:
\begin{enumerate}
    \item \textbf{Profitability}: mean annual returns ($MAR$).
    \item \textbf{Risk}: annualised volatility ($VOL$), downside deviation ($DDEV$), and maximum drawdown ($MDD$). $MDD$ depends on $VOL$, since higher leverage will also increase drawdowns. Therefore, we compare all strategies at a target volatility of $\sigma_{tgt} = 15\%$.
    \item \textbf{Risk-adjusted profitability}: annualised $Sharpe$, $Sortino$, and $Calmar$ ratios. 
    \item \textbf{Correlation ($CORR$)}: low (absolute) Pearson rank correlation coefficient to the Long-only benchmark helps produce consistent returns and avoid drawdowns during market crashes.
    \item \textbf{Transaction costs}: breakeven transaction cost ($BRK$) is expressed in basis points (bps), after optional volatility scaling.
    \begin{gather}
        BRK = \frac{\sum_{t=1}^{T}\sum_{i=1}^{N_A} \frac{w_{i,t}}{\sigma_{i,t}}\cdot r_{i,t+1}}{\sum_{t=1}^{T}\sum_{i=1}^{N_A}|\frac{w_{i,t}}{\sigma_{i,t}} - \frac{w_{i,t-1}}{\sigma_{i,t-1}}|}
    \end{gather}
\end{enumerate}

\subsection{Dataset Details}
\label{apdx:data}

We winsorise price data to 5 standard deviations, based on 252-day exponentially-weighted standard deviation. This is used to create features for DIN and benchmark models. We remove assets with more than 10\% missing data in the training set, since this negatively impacts the ability for convolutional layers to learn useful features. For the EUR50 dataset, we impose an additional constraint on selected assets to eliminate survivorship bias. Since DIN Feature Extractors use convolutional layers, the universe of assets within the training and test sets must be the same. Therefore, we fix assets in both training and test sets to be stocks that were members of EUR50 in the last week of the training set.

For the futures, cryptocurrency (Crypto) and FX datasets, the universe of assets is fixed. We list assets below and present further details in Exhibit \ref{table:Data_Sources}.
\begin{itemize}
    \item \textbf{Futures}: a list of contracts is available in \cite{wood2021trading}.
    \item \textbf{Crypto}: spot rates against USD for BCH, BTC, BTG, DASH, DOGE, ETC, ETH, LTC.
    \item \textbf{FX}: spot rates against USD for AUD, BRL, CAD, CHF, DKK, EUR, GBP, HKD, INR, JPY, KRW, MXN, NOK, NZD, SEK, SGD, THB, TWD, ZAR. 
\end{itemize}

\begin{table}[!h]
\centering
\caption{Dataset source, length and test sets.}
\begin{tabular}{||c c c c ||} 
 \hline
 \textbf{Futures} & \textbf{Equities} & \textbf{Crypto} & \textbf{FX}    \\
 \hline\hline
Pinnacle \cite{Pinnacle} & Compustat \cite{WRDS} & CMC \cite{CoinMarketCap} & FRB \cite{FRB}   \\ [0.5ex] 
 \hline
 Jan 2000 & Jan 2001 & Jan 2018 & Jan 2000\\
to Oct 2022 & to Dec 2022 & to Mar 2023 & to Jan 2023 \\[0.5ex] 
 \hline
 Test sets &  Test sets &  Test sets &  Test sets \\
2005 - 2010 & 2005 - 2010 & 2019 - 2020 & 2005 - 2010\\
2010 - 2015 & 2010 - 2015 & 2020 - 2021 & 2010 - 2015 \\
2015 - 2020 & 2015 - 2020 & 2021 - 2022 & 2015 - 2020 \\
2020 - 2022 & 2020 - 2022 & 2022 - 2023 & 2020 - 2023 \\[0.5ex] 
 \hline
\end{tabular}
\label{table:Data_Sources}
\end{table}

\subsection{Experiment Settings}
\label{apdx:experiment settings}

\begin{table*}[!b]
\centering
\caption{\centering{Hyperparameters for DIN Feature Extractors (FE). $hidden\_layer\_size$ is a parameter of the PS, but we choose lower values with AxialLOB FE to prevent out-of-memory (OOM) errors.}}
\begin{tabular}{||P{0.22\linewidth} || P{0.16\linewidth} | P{0.16\linewidth} | P{0.16\linewidth} | P{0.16\linewidth}||} 
 \hline
 Parameters & OrigCIM & FlexCIM  & DeepLOB & AxialLOB  \\
 
 \hline\hline
 hidden\_layer\_size, $N_H$ & [32, 64, 96, 128] & [32, 64, 96, 128] & [32, 64, 96, 128] & [16, 32, 48, 64] \\[0.5ex] 
 n\_filters & [16, 32, 48, 64] & [4, 8, 12, 16] & [16, 32, 48, 64] & [8, 16, 24]  \\ [0.5ex] 
 \hline
 ts\_filter\_length & [1, 3, 5, 10, 20] & - & - & - \\[0.5ex] 
 n\_filter\_layers & - & [1, 3, 5, 7, 9] & - & - \\[0.5ex] 
 n\_axial\_heads & - & - & - & [1, 2, 4] \\[0.5ex] 
 \hline
\end{tabular}
\label{table:hyperparam_choices_FE}
\end{table*}

Exhibits \ref{table:hyperparam_choices_FE}, \ref{table:fixed_params}, and \ref{table:tuning} show hyperparameter search grids and settings for tuning methods.

\begin{table}[!h]
\centering
\caption{\centering{Hyperparameters for LM and DIN PS.}}
\begin{tabular}{||c|| c | c ||} 
 \hline
 Parameters & LM & DIN (LSTM \& TFT) \\
 \hline\hline
 Objective & rank:pairwise & Eq. \ref{eqn:loss} \\[0.5ex] 
 Sequence length & - & 100 \\[0.5ex] 
 Max epochs & - & 250 \\[0.5ex] 
 Early stopping & 25 & 25 \\[0.5ex] 
 Train/valid ratio & 90\%/10\% & 90\%/10\% \\[0.5ex] 
 \hline
 Activation & - & ELU \\[0.5ex] 
 Normalisation & - & Layernorm \\[0.5ex] 
 $n\_heads$ & - & 4 (TFT only) \\[0.5ex] 
 $C_{valid}$ & - & 0  \\ [0.5ex] 
 $K_{valid}$ & - & 0   \\[0.5ex] 
 \hline
 $C$ & - & [0, 0.5, 1, 2, 5]    \\[0.5ex] 
 $K$ & - & [0, 1, 2, 5]    \\[0.5ex] 
 $dropout\_rate$ & - & [0.1, 0.2, 0.3]   \\[0.5ex] 
 $learning\_rate$ & [1e-6, 1e-5, 1e-4, & [1e-3, 1e-4, 1e-5] \\
     & 1e-3, 1e-2, 1e-1] &   \\[0.5ex]
 $n\_estimators$ & [50, 100, 250, & - \\
  & 500, 1000] &   \\[0.5ex] 
 $max\_depth$ & [6, 7, 8, 9, 10] & -   \\[0.5ex] 
 $reg\_alpha$ & [1e-6, 1e-5, 1e-4, & -  \\
  & 1e-3, 1e-2, 1e-1] &  \\[0.5ex] 
 $reg\_lambda$ & [1e-6, 1e-5, 1e-4, & -  \\
  & 1e-3, 1e-2, 1e-1] &  \\[0.5ex] 
  \hline
\end{tabular}
\label{table:fixed_params}
\end{table}

\begin{table}[!h]
\centering
\caption{\centering{Hyperparameter tuning settings.}}
\begin{tabular}{||P{0.37\linewidth} || P{0.13\linewidth} P{0.13\linewidth} P{0.13\linewidth} ||}
 \hline
 Parameters & HB & BO & RS  \\
 \hline\hline
 max\_epochs & 10 & -  & -\\ [0.5ex] 
 hyperband\_iterations & 1 & -  & - \\[0.5ex] 
 factor & 3 & - & -  \\[0.5ex] 
 \hline
 max\_trials & - & 25 & 50  \\[0.5ex] 
 num\_trial\_points & - & 20 & - \\[0.5ex] 
 alpha (noise) & - & 0.001 & - \\[0.5ex] 
 beta (exploration) & - & 10 & - \\[0.5ex] 
 \hline
 early\_stopping & - & 5 & 25 \\[0.5ex] 
 \hline
\end{tabular}
\label{table:tuning}
\end{table}

\subsection{Additional Results}
\label{apdx:additional_results}

\begin{table*}[!t]
\small
\centering
\caption{\centering{
    Comparison of DIN performance against benchmarks at $VOL = 15\%$. \textcolor{blue}{Blue} indicates that \textbf{not} using volatility scaling for individual assets improves portfolio performance and vice versa for \textcolor{red}{Red}. All benchmark strategies benefit from volatility scaling. Best values in each metric are \underline{underlined}. These are the same results as Exhibit \ref{table:Results_Pinnacle_NoCosts}, replacing $Sortino$, $Calmar$, $CORR$, $BRK$ with $PSR$, $MTR$, $HR$, $PNL$.}}
\begin{tabular}{@{\extracolsep{4pt}}lll|ccc|ccc|cc}
\toprule 
 DIN variant  & Tuning & Volatility & MAR & DDEV & MDD & Sharpe & PSR & MTR & HR & PNL  \\
 or benchmark & method & scaling & \% & \% & \% &  & \% & days & \% &   \\
  \midrule
 OrigCIM-TFT & HB & Yes & 44.5 & 9.1 & 21.6 & 2.53 & \textbf{100.0} & 215 & \underline{57.6} & 1.07 \\[0.5ex] 
 OrigCIM-TFT & HB & No & \textcolor{blue}{\underline{53.8}} & \textcolor{blue}{8.7} & \textcolor{blue}{13.8} & \textcolor{blue}{\underline{2.95}} & \textbf{100.0} & \textcolor{blue}{\underline{151}} & \textcolor{red}{57.4} & \textcolor{blue}{1.09} \\[0.5ex] 
 OrigCIM-TFT & BO & Yes & 48.0 & 9.2 & 18.9 & 2.69 & \textbf{100.0} & 205 & 57.4 & 1.08  \\[0.5ex] 
 OrigCIM-TFT & BO & No & \textcolor{blue}{49.8} & \textcolor{blue}{\underline{8.6}} & \textcolor{blue}{\underline{10.6}} & \textcolor{blue}{2.77} & \textbf{100.0} & \textcolor{blue}{168} & \textcolor{red}{56.1} & \textcolor{blue}{\underline{1.11}} \\[0.5ex] 
\midrule
 OrigCIM-LSTM & HB & Yes & 28.0 & 9.2 & 23.2 & 1.72 & \textbf{100.0} & 432 & 55.2 & 1.06 \\[0.5ex] 
 OrigCIM-LSTM & HB & No & \textcolor{blue}{30.4} & \textcolor{blue}{9.1} & \textcolor{red}{33.8} & \textcolor{blue}{1.85} & \textbf{100.0} & \textcolor{blue}{378} & \textcolor{red}{53.6} & \textcolor{blue}{1.09} \\[0.5ex] 
\midrule
JT LS & - & Yes & 9.2 & 10.7 & 25.9 & 0.66 & \textbf{99.7} & 3,166 & 52.0 & 1.02 \\[0.5ex]
LM LS & RS & Yes & 15.6 & 10.3 & 35.8 & 1.04 & \textbf{100.0} & 1,276 & 52.9 & 1.03\\[0.5ex]
MOP LO & - & Yes & 9.6 & 10.8 & 25.0 & 0.69 & \textbf{99.8} & 2,998 & 53.9 & 0.98 \\[0.5ex]
BAZ LO & - & Yes & 8.5 & 10.9 & 32.2 & 0.62 & \textbf{99.5} & 3,650 & 54.2 & 0.97 \\[0.5ex]
CMB EW & - & Yes & 12.4 & 10.8 & 20.7 & 0.86 & \textbf{100.0} & 1,938 & 54.9 & 0.98 \\[0.5ex]
CMB VS & - & Yes & 16.6 & 10.6 & 22.3 & 1.10 & \textbf{100.0} & 1,179 & 55.6 & 0.98 \\[0.5ex]
 \midrule
 Long-only & - & Yes & 4.4 & 10.9 & 41.6 & 0.37 & 93.8 & 10,340 & 52.8 & 0.98 \\[0.5ex] 
 \bottomrule
\end{tabular}
\label{table:Results_Pinnacle_HR_PNL}
\end{table*}

\begin{table*}[!b]
\small
\centering
\caption{\centering{ 
        Compare $PSR$ and $MTR$ for DIN variants on futures, equity, cryptocurrency, FX datasets. For comparability, all models use TFT Position Sizer (PS) and are scaled at portfolio level to $VOL = 15\%$. We choose a benchmark $Sharpe$ of 0 and significance threshold of 99\%. All results are with 0 costs.}}
\begin{tabular}{@{\extracolsep{4pt}}lll|cccc|cccc}
\toprule  
 DIN variant & Tuning & Volatility & \multicolumn{4}{c}{ Probabilistic Sharpe Ratio, $PSR$ $(\%)$ } & \multicolumn{4}{c}{ Minimum Track Record, $MTR$ (days) }   \\
  & method & scaling & Futures & Equity & Crypto & FX & Futures & Equity & Crypto & FX \\
 \midrule
 OrigCIM-TFT & HB & No & \textbf{100.0} & \textbf{100.0} & 55.7 & \textbf{100.0} & 151 & 1,032 & 373,100 & 798 \\[0.5ex]  
 FlexCIM-TFT & BO & No & \textbf{100.0} & \textbf{100.0} & 72.6 & \textbf{100.0} & 244 & 1,605 & 959 & 1,482 \\[0.5ex]
 DeepLOB-TFT & BO & No & \textbf{100.0} & \textbf{100.0} & \textbf{99.3} & \textbf{100.0} & 245 & 959 & 1,269 & 933 \\[0.5ex] 
 AxialLOB-TFT & HB & No & 93.6 & \textbf{100.0} & 40.4 & \textbf{100.0} & 1,049 & 1,482 & 130,900 & 977  \\[0.5ex] 
\midrule
 Long-only & - & No & 89.0 & 92.5 & 95.5 & 96.3 & 16,260 & 11,950 & 2,666 & 7,508 \\[0.5ex] 
 \bottomrule 
\end{tabular}
\label{table:Results_AllDatasets_FE_PSR}
\end{table*}

In Exhibit \ref{table:Results_Pinnacle_HR_PNL}, we present an extension to Exhibit \ref{table:Results_Pinnacle_NoCosts}, replacing $CORR$ and $BRK$ with hitrate $HR$ and profit-to-loss ratio $PNL$. By examining $HR$ and $PNL$, we can determine whether the strategy makes consistent, small profits or large, less frequent wins.
\begin{gather}
        HR = \frac{\mathrm{Count}(R_{positive})}{\mathrm{Count}(R_{p,t})}, \\ PNL = \frac{\mathrm{Mean}(R_{positive})}{\mathrm{Mean}(|R_{negative}|)},
\end{gather}
where transaction cost-adjusted portfolio returns $R_{p,t}$ are separated into positive $R_{positive} > 0$, negative $R_{negative} < 0$ and zero returns $R_{zero} = 0$. 

Exhibits \ref{table:Results_Pinnacle_HR_PNL} and \ref{table:Results_AllDatasets_FE_PSR} also extend Exhibits \ref{table:Results_Pinnacle_NoCosts} and \ref{table:Results_AllDatasets_FE_SharpeMDD} by examining the statistical significance of strategy $Sharpe$ compared to a benchmark $Sharpe$ of 0. We interpret the Probabilistic Sharpe Ratio ($PSR$) \cite{bailey2012sharpe} as significant when above a 99\% confidence level, and highlight this in \textbf{Bold}. The Minimum Track Record ($MTR$) is the minimum number of observations, in days, needed for a significant result.

\end{document}